\documentclass[preprint,12pt]{elsarticle}
\usepackage{lineno,hyperref}
\journal{Journal of Systems and Software}
\usepackage{fancyhdr}
\fancypagestyle{pprintTitle}{%
  \fancyhf{}%
  \fancyfoot[C]{\small\textit{Preprint submitted to Journal of Systems and Software, June 2026}}%
}
\newlength\MAX  
\usepackage{booktabs}
\usepackage{subfig}
\usepackage{needspace}
\usepackage[framemethod=tikz]{mdframed}

\mdfdefinestyle{mpdframe}{
    frametitlebackgroundcolor   =black!15,
    frametitlerule              =true,
    roundcorner                 =5pt,
    middlelinewidth             =1pt,
    innermargin                 =0.3cm,
    outermargin                 =0.3cm,
    innerleftmargin             =0.3cm,
    innerrightmargin            =0.3cm,
    innertopmargin              =0.5cm,
    innerbottommargin           =0.5cm,
}

\usepackage{lipsum}
\usepackage[most]{tcolorbox}

\tcbset{
  mpd/.style={
    enhanced,
    breakable,
    colback=black!5,
    colframe=black!70,
    colbacktitle=black!15,
    coltitle=black!80,
    title filled,
    fonttitle=\bfseries,
    boxrule=0.8pt,
    arc=5pt,
    left=3mm,right=3mm,top=4mm,bottom=4mm
  }
}

\usepackage{xcolor}
\newcommand{\nb}[1]{#1}

\newcommand{\eg}{\textit{e}.\textit{g}., }

\usepackage{natbib}
\usepackage{soul}\setuldepth{article}
\usepackage{soulutf8}

\begin{document}

\begin{frontmatter}

\title{The Perception and Impact of Non-inclusive Language in Software Artifacts}

\newif\ifblindreview
\blindreviewfalse    

\ifblindreview
  \author[inst1]{Anonymized for review}
  \address[inst1]{\mbox{}}
\else
  \author[mymainaddress]{Ahmad J. Tayeb\corref{mycorrespondingauthor}}
  \cortext[mycorrespondingauthor]{Ahmad J. Tayeb}
  \ead{ajtayeb@kau.edu.sa}
  \author[mysecondaryaddress]{Mohammad D. Alahmadi}
  \ead{mdalahmadi@uj.edu.sa}
  \address[mymainaddress]{Department of Information Technology, Faculty of Computing and Information Technology, King Abdulaziz University, Jeddah, Saudi Arabia}
  \address[mysecondaryaddress]{Department of Software Engineering, College of Computer Science and Engineering, University of Jeddah, Jeddah, Saudi Arabia}
\fi

\begin{abstract}
Terminology such as ``whitelist/blacklist,'' ``master/slave,'' ``man-hours,'' or ``dummy value'' has long been part of the technical vocabulary used in software artifacts, including source code, version histories, and documentation. In recent years, however, many of these expressions have been recognized as potentially non-inclusive and unwelcoming to groups historically underrepresented in software development, such as people of color, women, and individuals with disabilities. Consequently, a growing movement within the software industry has sought to replace these terms with more inclusive alternatives. Despite these initiatives, little is empirically known about how software developers perceive such terminology or how its continued use may influence their professional experiences and sense of belonging.

This paper addresses the knowledge gap by examining how software developers perceive non-inclusive terminology in software and its perceived impact on team dynamics, productivity, belonging, and well-being. We surveyed open-source contributors and received 1,517 responses, of which 1,212 were complete and analyzed. On average, respondents reported low negative workplace impact overall; however, perceptions and impacts varied by demographic group. \nb{Women and non-binary participants, as well as respondents residing in the United States, were more likely to view the terms as non-inclusive. Among those who considered the terminology non-inclusive, non-binary participants reported higher overall negative impacts than male respondents, and female participants reported higher impact specifically on their sense of belonging.}
\end{abstract}


\end{frontmatter}

\section{Introduction}\label{sec:introduction}

Team diversity in the workplace has been associated with positive outcomes such as improved problem solving, better product design, and greater creativity and innovation, especially when the tasks addressed are complex and the individuals in the team possess a wide range of knowledge, skills, experiences, and backgrounds \citep{bantel1989top,ancona1992demography, pelled1997demographic,williams1998reilly}. In software development, diverse teams have been linked to increased team performance and productivity \citep{vasilescu2015gender,ortu2017diverse}. \nb{However, the relationship is nuanced: while knowledge diversity has been associated with performance benefits through constructive task conflict, value diversity may increase relationship conflict and adversely affect team performance~\citep{ liang2007effect}.} 

\nb{Nonetheless}, the software development field is known to struggle with a lack of diversity in its labor force in many countries around the world. For example, in 2021, in the European Union, women made up only 19\% of the employees in information and communication technology, including software development \citep{eurostat_2022}. In the United States, data from the US Bureau of Labor Statistics \citep{us_bureau_labor_statistics_2021} show that in 2021, only 20\% of software developers and computer programmers were women, 5.5\% were Black or African American, and 6\% were Hispanic or Latino. This is in sharp contrast to the percentages of these groups in the general population of the US, where women, Black or African American, and Hispanic and Latino people represent 51\%, 14\%, and 19\% of the population, respectively \citep{us_census_2021}. In contrast, 57\% of software developers and computer programmers were white (a.k.a.\ Caucasian, of European descent) in 2021, which is consistent with the 59\% of the general US population they represent. \nb{While these figures reflect representation gaps rather than direct evidence of inequitable hiring, they motivate ongoing efforts to understand and address barriers to diversity and inclusion in software development.}

\nb{At the same time, open source software development faces the same diversity challenges.} For example, women represent less than 10\% of code contributors \citep{beecham2008motivation} and are likely to disengage from participation on GitHub earlier than men \citep{qiu2019going}. Although efforts have been made to increase diversity, equity, and inclusion, especially in large companies and open source communities such as Microsoft, Google, and Mozilla \citep{blincoe2019perceptions}, there is still a long way to go before achieving a software developer population that reflects the demographics of the general population.

A recent area of effort to promote software developer diversity and inclusion is to discontinue the use of non-inclusive language in software artifacts, such as `master/slave', `blacklist/whitelist', `man-hours', `dummy value', etc. While these terms have been commonplace in the technical vocabulary of software developers and used in software artifacts such as code, version history, and documentation for decades, they have also been associated with negative connotations tied to racism, sexism, \nb{and other forms of discrimination}. One of the first instances of non-inclusive terminology deprecation occurred in 2014 when Drupal, a web framework, transitioned to using `primary/replica' instead of `master/slave' \citep{drupal_2014}. The Django web framework also deprecated the use of `master/slave', instead using `leader/follower' \citep{django_2014}. The Python programming language fully deprecated the `master/slave' feature from its code base in 2018 \citep{python_2018}. More recently, in 2020, Twitter Engineering announced its commitment to transition away from using terms it identified as non-inclusive in its software artifacts, accelerated by the Black Lives Matter Movement and lobbying by black engineers at the company \citep{shankland_2020}. Other similar announcements were made in 2020 by companies and projects such as the Linux Foundation, GitHub, LinkedIn, Jenkins, Apple, Android, Google Chrome, and Go \citep{hunter_2020}. In 2021, the Inclusive Naming Initiative was born, which is a cross-organizational effort to remove harmful, racist, and unclear language from software projects. Its sponsors include Intel, Cisco, RedHat, and the Continuous Delivery Foundation \citep{inclusive_naming_initiative_2021}.

These initiatives have led to heated online discussions among developers about whether such terminology has an impact on diversity and inclusion and whether it is worthwhile for developers to adopt alternative terminology \citep{cosset_2020}. However, no research study has yet investigated how the use of non-inclusive terminology in software is perceived by or how it impacts software developers, including those from groups the terminology is considered to directly affect.

In this paper, we aim to understand how software developers perceive the use of non-inclusive terminology in software and how it may affect their professional experiences, team interactions, and overall well-being. To address this goal, we conducted an anonymous online survey of active open-source software contributors. The terminology considered in our study is based primarily on the list released by X (formerly Twitter) Engineering \citep{shankland_2020}, which, at the time of data collection, represented one of the most comprehensive and widely referenced sets of terminology identified for replacement across the software industry. In total, we received 1,212 complete responses.

Our analysis showed that, overall, most demographic groups, including some that might be directly affected by non-inclusive terminology, reported limited negative workplace impact from such language. However, perceptions varied across demographics. A higher proportion of women, non-binary, older, and U.S.-based participants viewed the terminology as non-inclusive, while respondents identifying as White, Hispanic, or Asian tended to be more neutral. Perceptions also differed across individual terms, with some considered less problematic than others across all groups.

Many participants who did not view the terminology as harmful emphasized that meaning depends on context and that, within software development, these terms are often interpreted technically rather than socially. In contrast, over half of those who regarded the terminology as non-inclusive felt that its continued use negatively affected their feelings of inclusion and the diversity of their teams.

\section{Methodology}\label{sec:methodology}
In this section, we first describe the terminology considered in this study, the non-inclusive connotations associated with it, and the more inclusive alternatives generally proposed to replace it. Then, we describe the design of our study, including the survey we distributed to software developers, the means of its distribution, and the research questions we aim to address.


\subsection{Terminology}
\label{sec:Terminology}

The terminology we focus on in this study is inspired by a tweet posted by X (formerly Twitter) Engineering on their official account on July \nb{2020} \footnote{https://twitter.com/TwitterEng/status/1278733305190342656?s=20}. The tweet contains a list of terms considered non-inclusive, along with more inclusive alternatives for each term. Furthermore, it states the commitment of the X Engineering team to gradually replace the non-inclusive terms with alternatives in all their software artifacts, including source code and documentation. This list of terms, depicted in \hyperref[tab:twitter_terminology]{Table}~\ref{tab:twitter_terminology}, represents one of the most comprehensive lists of terminology on this topic that we were able to identify by the beginning of this project. Since then, other lists have become available, such as the one adopted by the Inclusive Naming Initiative \citep{inclusive_naming_initiative_2021}, which includes some, but not all, the terms we use in our study while also adding others. We plan to investigate those additional terms in future work. 

\begin{table}[h]
  \caption{X (formerly Twitter) engineering terminology changes}
  \label{tab:twitter_terminology}
  \centering
  \resizebox{\columnwidth}{!}{
  \begin{tabular}{ll}
  \toprule
  \multicolumn{1}{l}{\textbf{Non-inclusive terms}} & \multicolumn{1}{l}{\textbf{Preferred inclusive alternatives}} \\
  \midrule
  Whitelist                             & Allowlist                                           \\
  Blacklist                             & Denylist                                            \\
  Master/slave                          & Leader/follower, primary/replica, primary/standby   \\
  Grandfathered                         & Legacy status                                       \\
  Gendered nouns (e.g., ``guys'')       & Folks, people, you all, y’all                       \\
  Gendered pronouns (e.g., ``he/him/his'') & They/them/their                                   \\
  Man hours                             & Person hours/engineer hours                         \\
  Sanity check                          & Quick check, confidence check, coherence check      \\
  Dummy value                           & Placeholder value, sample value                     \\
  \bottomrule
  \end{tabular}
  }
\end{table}

\textbf{Whitelist/blacklist:} Whitelist and blacklist refer to lists of users, devices, or features that are allowed or blocked, respectively, within a system. The colors white and black do not have any non-inclusive meaning by themselves; the non-inclusive connotation is given to them in this context, where white is synonymous with ``good,'' and black is synonymous with ``bad.'' Proponents of deprecating this terminology state that it applies metaphor and unintended meaning where it is not needed \citep{inclusive_naming_initiative_2021}. The alternatives allowlist/denylist are considered more inclusive as well as more descriptive. Companies that replaced whitelist and blacklist with more inclusive alternatives include Delphix, X (formerly Twitter), Go, Google Chromium, Android, and PagerDuty.

\textbf{Master/slave:} The words master and slave have been widely used for decades in software development and usually refer to a control relationship between processes or devices, where the ``master'' device has control over or delegates work to ``slave'' devices \citep{eglash_2007}. Sometimes master is used to denoting something that leads, serves as a primary resource, or is considered first. Since 1976, the US has issued more than 67,000 patents using the terms master and slave \citep{landau_2020}.

The terms master and slave, however, imply connotations of ownership and subjugation. Proponents of deprecating this terminology state that ``master-slave is an oppressive metaphor that will and should never become fully detached from history,'' and that the terms are not welcoming to people of color whose ancestors were subjected to inhumane practices this terminology originates from \citep{inclusive_naming_initiative_2021}. Companies and projects that have removed master and slave from their software artifacts include GitHub, X (formerly Twitter), Python, Django, Drupal, Go, Android, Jenkins, Delphix, and PagerDuty.

\textbf{Grandfathered:} The term grandfathered (sometimes used as ``grandfathered in'') refers to allowing a preexisting use or service to continue despite later regulation \citep{moore_1996}. The use of the term relates back to the ``grandfather clause'' introduced in the United States in the late 19th century by some Southern states, which passed legislation that made it extremely hard for African-Americans to vote. The laws introduced new requirements for voters to pass literacy tests, pay poll taxes, and have residency/property ownership. States would exempt citizens from these new requirements if their grandfathers had voting rights before the Civil War. The term grandfathered arose from this clause. The motivation and outcome of these laws were to keep poor and illiterate former slaves and their descendants from voting while allowing poor and illiterate whites to vote \citep{greenblatt_2013}.
Similar to the terms master and slave, the usage of grandfathered invokes connotations that can be unwelcoming to people of color and provide unintentional meaning. Proponents for replacing this terminology prefer the usage of legacy or legacy status, as it implies the intent of pre-existing conditions exempting someone from new requirements, without the harmful connotations of the grandfather clause \citep{riley_2019}. Companies that replaced grandfathered with this alternative include X (formerly Twitter) and HubSpot.

\textbf{Gendered nouns and pronouns:} In English, pronouns are used to refer to a noun or individual in place of the noun or individual’s name. Gendered pronouns refer to a person’s gender, while non-gendered pronouns are not tied to a specific gender and are used for those who identify as non-binary\citep{college_2020}. It is common in general conversation to assume the gender of others through their name or appearance and, in turn, use gendered language based on these assumptions. Also, it is common to use male-gendered nouns like guys to address a group of people. However, using the wrong gender nouns and pronouns can lead to women, trans, and non-binary individuals being misgendered, which may cause them to feel marginalized \citep{college_2020}. It is recommended to use the gender-neutral pronouns they/them and gender-neutral nouns like folks, people, and you all until gender information can be obtained \citep{tobia_2016}. Projects that made their code base = gender-neutral by replacing gendered nouns and pronouns include Google Chromium and X (formerly Twitter).

\textbf{Man hours:} A man hour refers to a unit of one hour’s work by one individual. However, similar to gendered nouns and pronouns discussed above, man hours explicitly refers to the male gender, which may cause female or non-binary individuals to feel that men’s work hours are more highly regarded. Suggestions for replacement mirror gendered nouns by using a gender-neutral term such as staff hours.

\textbf{Sanity check:} A sanity check refers to quickly validating that a claim or calculation is correct. The usage of sanity in this context is considered to conflate mental illness with a relatively quick and trivial analysis. Proponents for replacing this phrase in software suggest using a more neutral phrase that more directly relates to the task at hand, such as coherence check \citep{hansen_2017,google_2022b}.

\textbf{Dummy value:} A dummy value in software generally refers to a value used as a temporary placeholder. Terms such as ”dummy” or ”moron” have historically been used to categorize people with mental disabilities as lesser humans \citep{pwda_2022}. The usage of dummy in this context contributes to this stigma since a dummy value is an insignificant part of the software. Proponents for replacing this phrase suggest using an alternative that references the task of the value more directly, such as the literal placeholder value \citep{google_2022a}.


\subsection{Survey Design}
\label{sec:Survey_Design}

To study the perception of software developers regarding the hypothesis that certain software terminology has non-inclusive connotations and implications in software, we designed an anonymous online survey composed of five main sections as follows. The first section was the informed consent for study participation, which introduced the survey goals and scope, the anticipated survey duration (10-15 min), the anonymity of the responses, and listed the contact information of the researchers and the human research review committee, which reviewed and approved the study.

The second section of the survey contained questions aimed at understanding the demographic makeup and background of the study participants, namely, questions regarding their primary occupation, programming experience, gender, age, race, country of origin, country they currently reside in, and whether English is their native language or not.

The third section of the survey introduced the list of non-inclusive terms we considered in the study (see Section \ref{sec:Terminology}) and asked participants whether they agreed with the hypothesis that the terms are non-inclusive. This hypothesis is inspired by the industry initiatives aimed at replacing these terms. The respondents could answer using a five-point Likert scale: Strongly Disagree (1), Disagree (2), Neither Agree nor Disagree (3), Agree (4), and Strongly Agree (5). When participants disagreed with the hypothesis, we were also interested in knowing their reasoning. We listed four potential reasons and included the option for participants to write in an answer. The four listed reasons were: “(1) My native language/(2) My culture does not share the negative connotations of the terms”, “(3) I do not believe that these terms have an impact in the context of software”, “(4) These terms do not have a personal impact on me or my associates”. We concluded the section with an open question about any other non-inclusive language they encountered in software artifacts or developer communications. 

The fourth section of the survey included questions aimed at understanding which of the non-inclusive terms we considered are used by our participants or their teams in software projects and whether they use any of the alternative terminology meant to replace the non-inclusive terms in their software projects. 

The last survey section asked participants about the impact of using non-inclusive terms in software on: (i) the diversity of their team, (ii) the inclusion and acceptance of diverse ideas at work, (iii) their self-esteem, (iv) their work productivity, and (v) their sense of belonging in an organization. Participants could answer using the same 5-point Likert scale mentioned above. For the participants who indicated that the use of the terms considered in this study did not have any impact on one of the listed aspects, we were also interested to know why that was the case. We, therefore, asked them to provide us with a rationale and offered five options, plus the option to write their own response. The five options were the same as in the third section of the survey, plus an additional one: “My company’s culture does not share the negative associations of the terms.” This section concluded with an open question asking participants to list any other ways non-inclusive language has negatively impacted their life and work environment. Finally, we gave participants the opportunity to share any final thoughts they had regarding non-inclusive language in software artifacts and developer communications. The complete survey is available in our online replication package \citep{replication_package}. While the results we present in Section \ref{sec:results} represent the majority of our findings, we were not able to include all results due to space limitations. However, our replication package \citep{replication_package} contains the complete results, as well as the raw responses we collected.

\subsection{Recruitment and Sampling}
\label{sec:Survey_Distribution}

Our target population was active software contributors to open-source projects. We used purposive sampling to reach developers who had demonstrable contribution history on GitHub. Recruitment invitations were sent to contributors via publicly available professional contact channels. Each invitation described the study, emphasized voluntariness and anonymity, and linked to the survey.

To include a diverse set of projects, we sampled projects of varying popularity and domains and contacted a subset of contributors from each, avoiding duplicate outreach. We did not collect or store personal identifiers beyond what was necessary to send the invitation and to enforce one response per invite; identifiers were not linked to survey responses.

\textbf{Ethical considerations:}
The study protocol, consent language, and recruitment approach were reviewed and approved by our institutional ethics board. We limited contact to information that contributors had chosen to make public for professional purposes, used it solely for a one-time research invitation, and honored non-participation and opt-out. No incentives were offered. Data processing followed data minimization and purpose-limitation principles; contact records were deleted after recruitment and were never stored with response data.

\subsection{Data and Analysis}
\label{sec:Survey_Data_Analysis}

Data collection spanned two months. We received 1{,}517 submissions; after excluding incomplete responses per a pre-registered rule (missing mandatory sections), 1{,}212 complete responses remained for analysis. Median completion time was 7 minutes.

We used descriptive statistics for central tendencies and distributions of Likert items, and stratified summaries by demographic groups (e.g., age, gender, region, native vs. non-native English). For exploratory patterning, we experimented with simple decision-tree models (CART) to surface interaction effects; due to class imbalance and overfitting risk, we report only robust descriptive findings in the main text and include exploratory artifacts in the replication package. \nb{Two authors independently coded the responses and subsequently met to resolve disagreements through discussion until consensus was reached. All disputed cases were resolved collaboratively, and the final categories were validated by both authors.} \nb{Where multiple primary statistical tests are conducted within a single research question, we apply Holm--Bonferroni correction to control the family-wise error rate, treating all tests per RQ as one comparison family; the complete set of corrected thresholds is reported in our replication package \citep{replication_package}.}

All de-identified data, figures, and analysis scripts are provided in the replication package \citep{replication_package}.


\subsection{Research Questions}
\label{sec:Research_Questions}

The goal of our study is to understand the perceptions of software contributors on the use and impact of non-inclusive terminology in software. With this goal in mind, we formulated \nb{two} research questions, which we aim to answer: 

\begin{itemize}
\item{\textbf{RQ1. What is the perception of software contributors on the non-inclusive connotations of the studied terminology in software?}}
This question is motivated by the need to shed light on software contributors' attitudes and beliefs toward using non-inclusive terminology in software development and its potential impact on their work and the broader community.
\item{\textbf{RQ2. What is the perceived impact of the use of non-inclusive terminology in software on the lives and work environments of software contributors?}}
The motivation for this question is to understand the real-world consequences of the use of non-inclusive terminology and how this may impact the experiences of software contributors in their daily work.
\end{itemize}

As the movement to replace non-inclusive terms with inclusive ones originated in U.S.-based institutions, we acknowledge the potential for cultural differences in software contributors' perspectives. To explore these differences, we will distinguish between U.S.-based and non-U.S.-based participants in our data analysis. Therefore, for our study, we define U.S. participants as those residing in or originating from the United States, while non-U.S. participants refer to those not residing in and not originating from the United States. By addressing these research questions, our study aims to provide insights into the impact of non-inclusive terminology use in software development on software contributors' work and lives.

\section{Results}\label{sec:results}

\nb{In addition to descriptive statistics, we performed non-parametric statistical tests (Mann–Whitney U for two groups and Kruskal–Wallis for multiple groups) to assess whether observed differences between demographic groups are statistically significant. Given the ordinal nature of Likert-scale data, these tests are more appropriate than parametric alternatives. We report statistically significant differences where applicable and note when observed differences are not statistically significant or have small effect sizes. Multiple-comparison correction is applied per family (one family per RQ) as described in Section~\ref{sec:Survey_Data_Analysis}; findings that survive Holm--Bonferroni correction are noted in each summary box.}
\subsection{Participant Demographics}
\label{sec:Participant_Demographics}

In this section, we briefly describe the demographics of our study participants (1,212) based on different characteristics, in order to put our results into context.

\begin{figure}[htbp]
    \centering
    \subfloat[\centering Age]{{\includegraphics[width=0.45\linewidth]{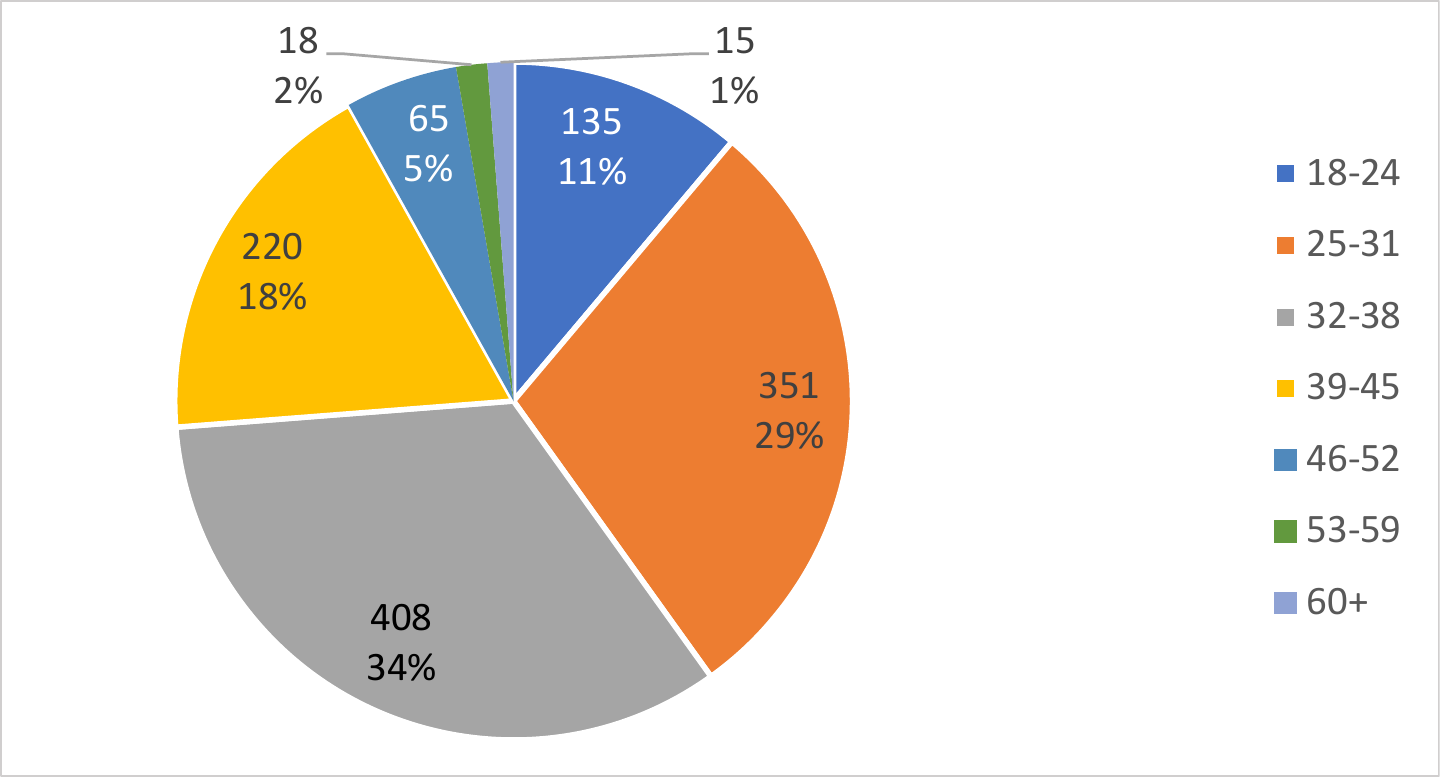} }}%
    \centering
    \subfloat[\centering Gender ]{{\includegraphics[width=0.45\linewidth]{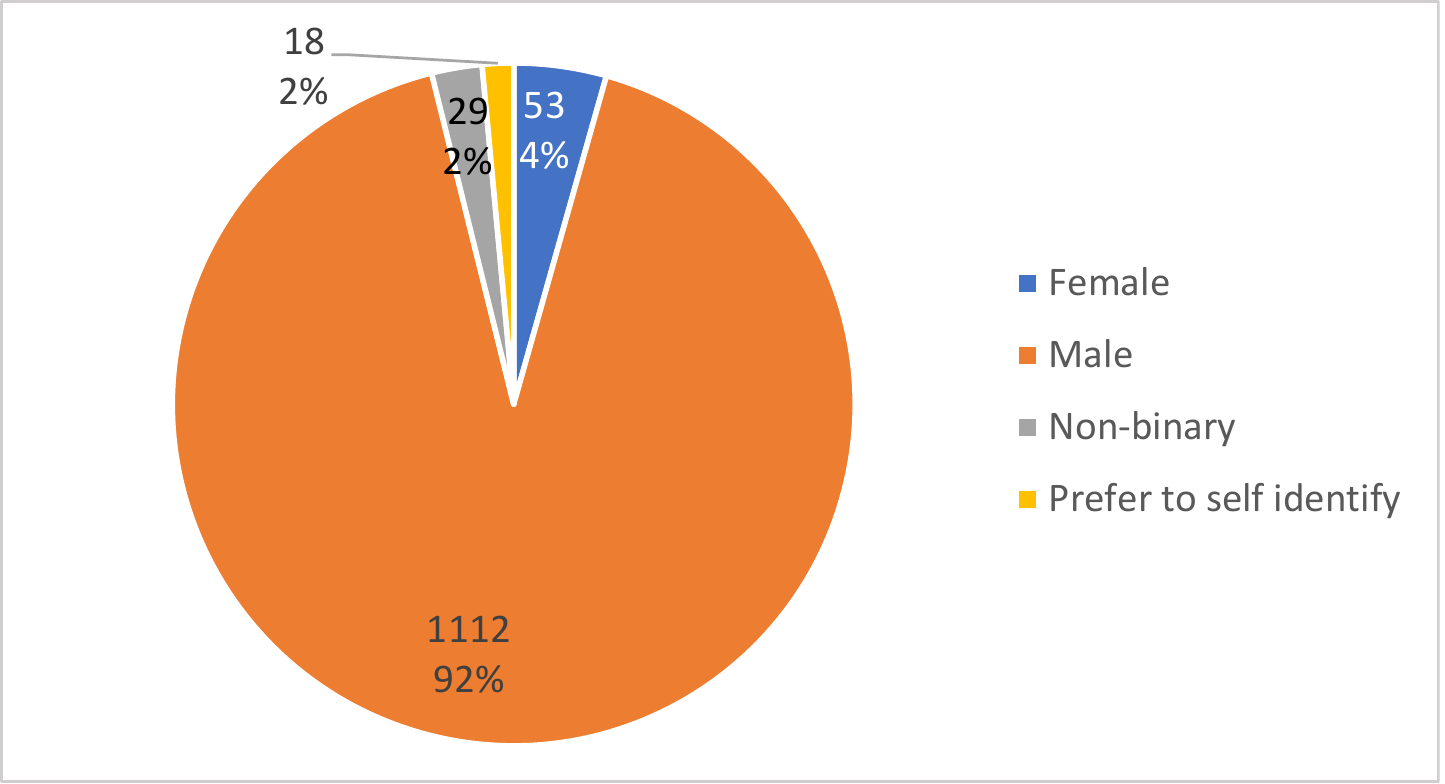} }}%
    \caption{Demographic distribution of respondents by (a) age and (b) gender}%
    \label{fig:q0_age_gender}%
\end{figure}

\textbf{Age:} The age distribution of the study participants is presented in Figure \ref{fig:q0_age_gender}a. Our results indicate that the vast majority of respondents fall within the age range of 25-45 years, comprising 81\% of the sample. A smaller proportion of respondents are less than 25 years old (11\%), while only \nb{8}\% are older than 45. The data suggests that our study population primarily comprises individuals from the Millennial generation. 

\textbf{Gender:} The gender distribution of our study participants is as follows: 92\% identified as male, 4\% identified as female, and 4\% identified as non-binary or preferred to self-identify (Figure \ref{fig:q0_age_gender}b). While the representation of non-male genders in our sample is low, this finding is consistent with the gender distribution reported in other developer surveys. For example, the 2021 Stack Overflow Software Developer Survey \citep{stack_overflow_2021} reported that 92\% of respondents identified as male, while a 2017 GitHub survey \citep{zlotnick_2017} found that 95\% of respondents were male. 

\begin{figure}[htbp]
    \centering
    \subfloat[\centering Primary occupation]{{\includegraphics[width=0.45\linewidth]{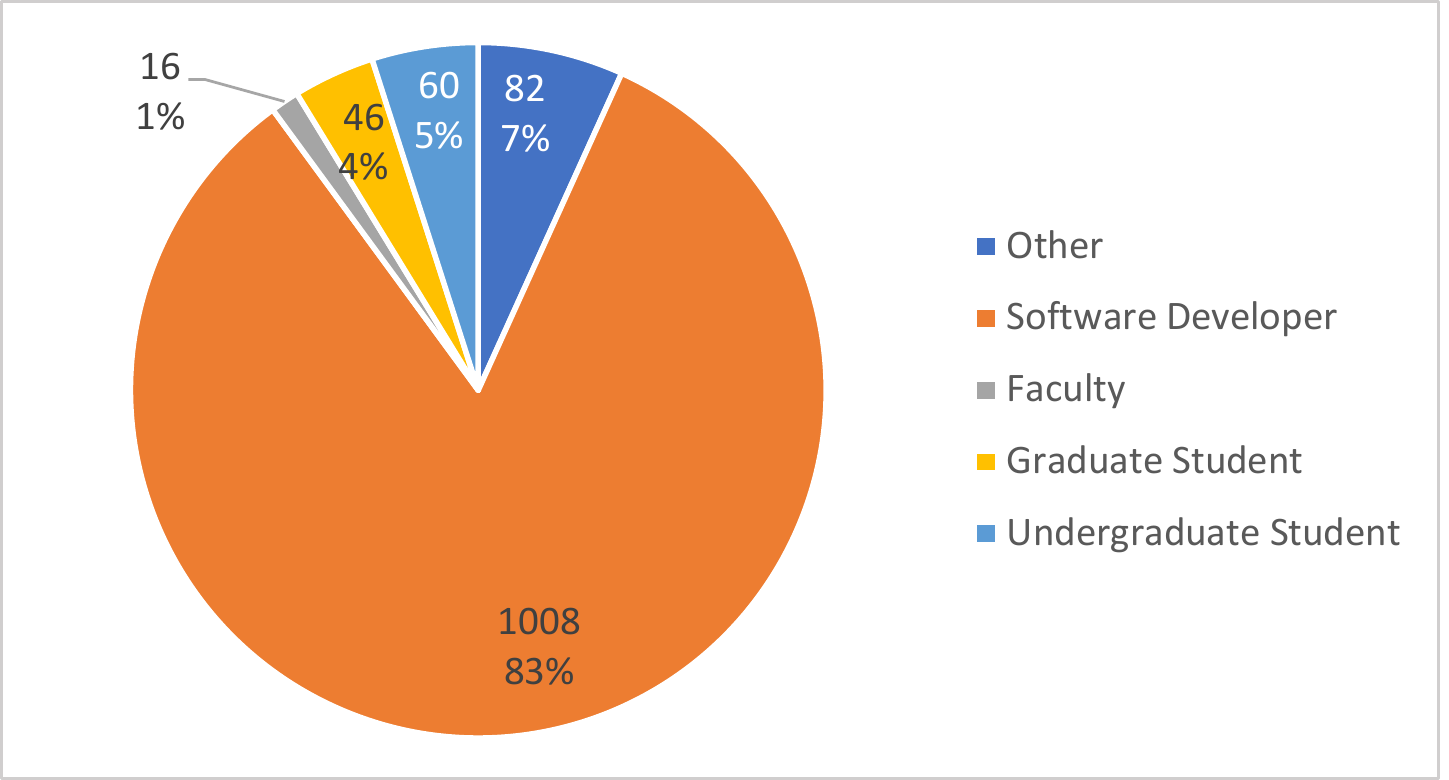} }}%
    \centering
    \subfloat[\centering Programming experience  ]{{\includegraphics[width=0.45\linewidth]{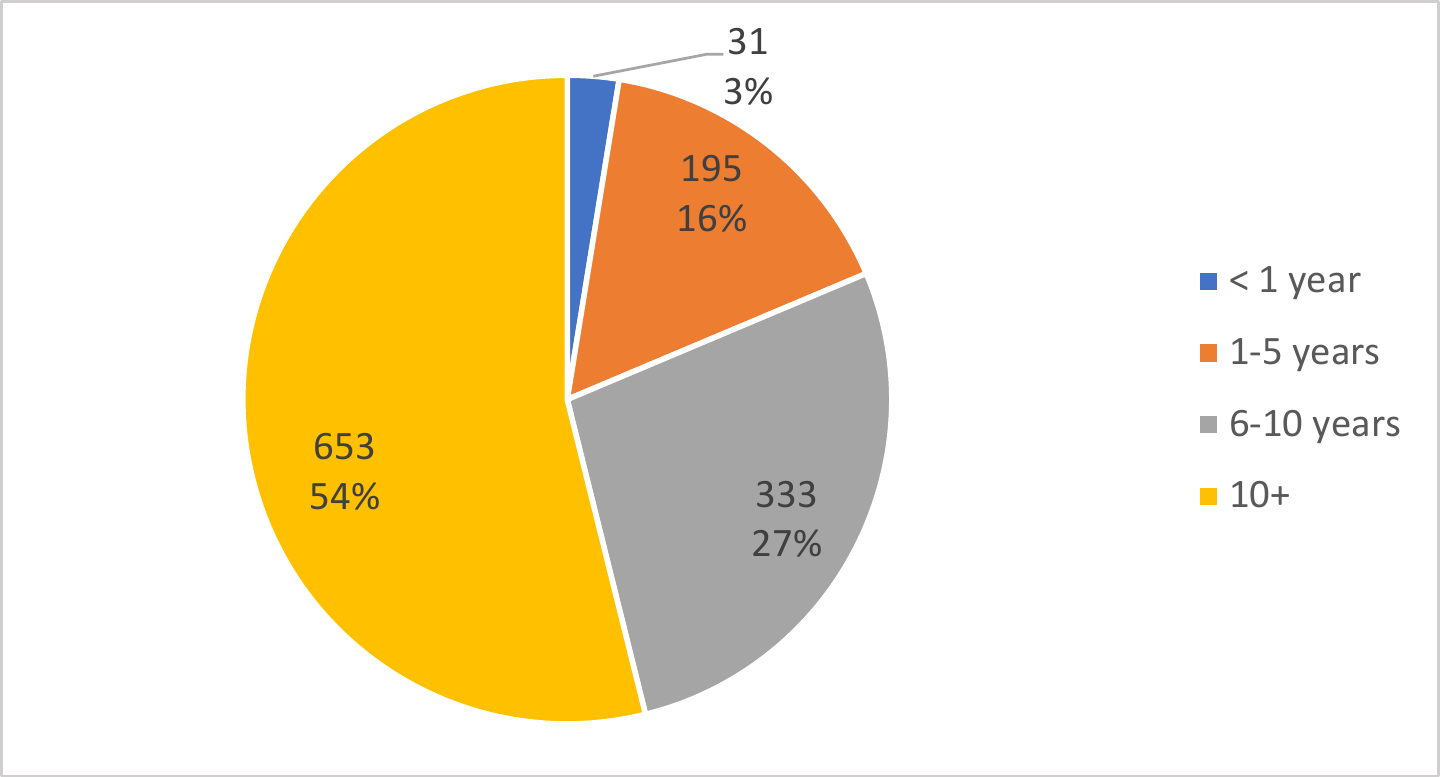} }}%
    \caption{Demographic distribution of respondents by (a) primary occupation and (b) programming experience}%
    \label{fig:q0_experience}%
\end{figure}

\textbf{Occupation:} The majority of respondents to our survey are professional software developers, representing 83\%. In comparison, 5\% of the respondents are undergraduate students, 4\% are graduate students, and 1\% are faculty (Figure \ref{fig:q0_experience}a). Finally, 7\% of the respondents have other types of occupations. Therefore, we can conclude that the responses to our survey reflect mostly the opinion of professional software developers.

\textbf{Programming experience:} The distribution of programming experience among our study participants is shown in Figure \ref{fig:q0_experience}b. Our results indicate that 81\% of respondents have more than six years of programming experience, while 19\% have five years or less. Therefore, we can conclude that the vast majority of our participants are experienced developers.

\textbf{Race:} The majority of our respondents are White (72\%), followed by Asian/Pacific Islanders (13\%) and Hispanic or Latinx participants (6\%). Only 1\% of our respondents were Black or African American and 1\% were Native American. The remaining 7\% identified themselves as belonging to other races. 

\begin{figure}[htbp]
    \centering
    \subfloat[\centering Continent of origin]{{\includegraphics[width=0.45\linewidth]{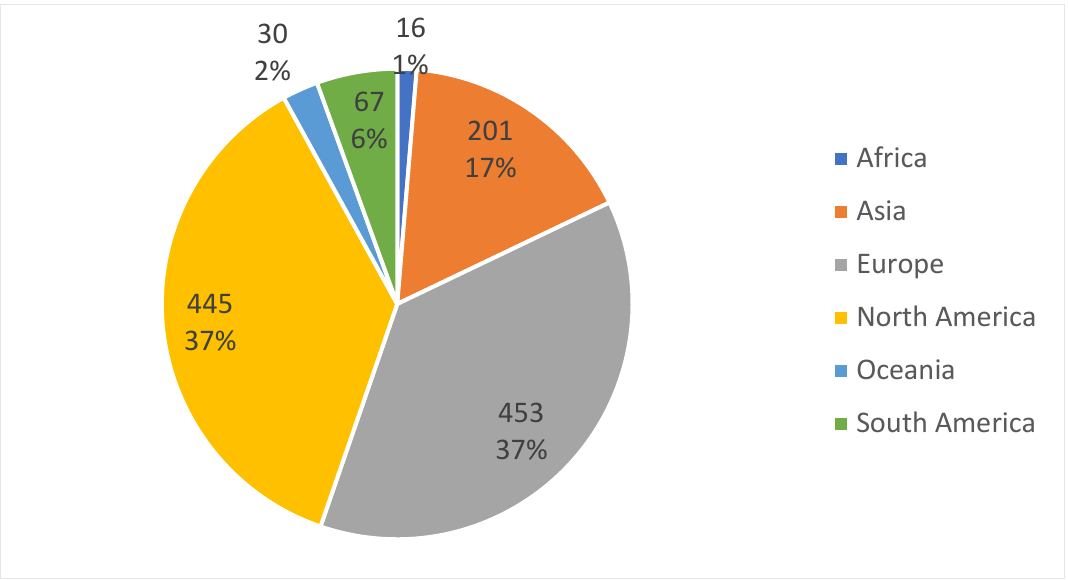} }}%
    \centering
    \subfloat[\centering Continent of residence]{{\includegraphics[width=0.45\linewidth]{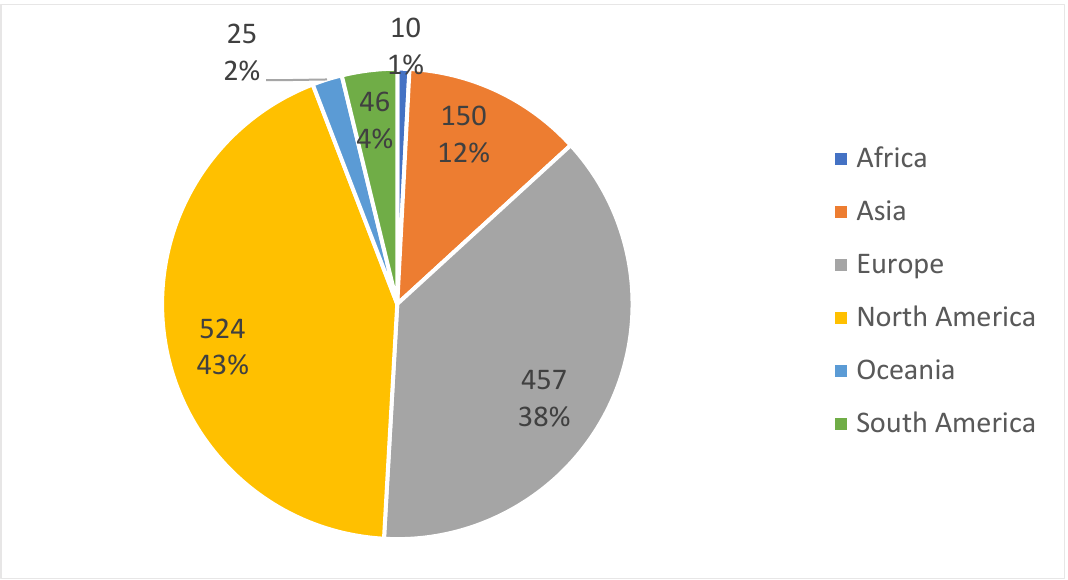} }}%
    \caption{Geographic distribution of respondents by continent of origin (a) and residence (b)}%
    \label{fig:q0_continent}%
\end{figure}

\textbf{Continent of origin:} The data presented in Figure \ref{fig:q0_continent}(a) indicates that most respondents are originally from North America (37\%) and Europe (37\%). Next, participants originally from Asia make up 17\%. Finally, the remaining participants are originally from South America (6\%), Oceania (2\%), and Africa (1\%).

\textbf{Continent of residence}: Figure \ref{fig:q0_continent}b shows the distribution of our survey respondents by continent of residence. The majority of respondents (43\%) reside in North America, followed by Europe (38\%). The remaining respondents are distributed across Asia (12\%), South America (4\%), Oceania (2\%), and Africa (1\%). These proportions are similar to those observed in the origin information of the respondents (Figure \ref{fig:q0_continent}a).

\begin{figure}[htbp]
\centering
\includegraphics[width=0.75\linewidth]{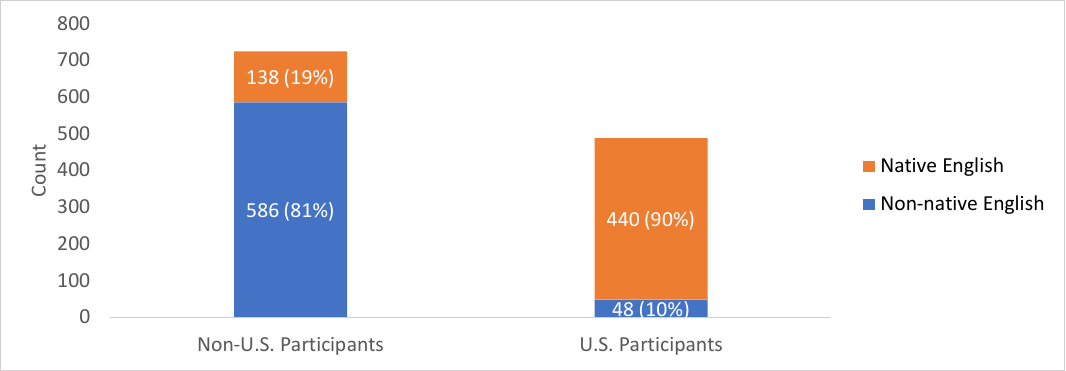}
\caption{The distribution of participants who are native and non-native English speakers in US and other countries.}
\label{fig:q0_english}
\end{figure}

\textbf{English as native language:} Overall, 52\% of the respondents are non-native English speakers. Moreover, 80\% of native English-speaking participants originate from or reside in the United States. In contrast, 92\% of non-native English-speaking participants are not originally from or residing in the U.S. (U.S.-based participants), as shown in Figure \ref{fig:q0_english}. Also, 90\% of the U.S.-based participants in our study are native English speakers, while only 19\% of non-U.S. participants are native English speakers.

\vspace{2cm}

\begin{tcolorbox}[mpd, title={Summary (Participant Demographics)}]
Our study analyzed 1,212 valid responses. Most respondents were between the ages of 25 and 45 (81\%) and identified as male (92\%). The majority were professional software developers (83\%) with over six years of programming experience (81\%). Most participants identified as White (72\%) and were primarily from North America or Europe. Additionally, 52\% of respondents were non-native English speakers, while most native English speakers were from or residing in the United States (72\%). These characteristics provide context for interpreting the study’s findings and indicate that the participant pool largely consisted of experienced, male, White, English-speaking software professionals based in North America or Europe.

\nb{Our sample is broadly consistent with prior large-scale developer surveys. For example, the gender distribution (92\% male) closely aligns with the Stack Overflow Developer Survey, which consistently reports that over 90\% of respondents identify as male} \cite{stack_overflow_2021,stackoverflow_survey_kaggle_2024}. \nb{Similarly, the predominance of contributors from North America and Europe reflects trends observed in GitHub’s Octoverse reports} \cite{github_octoverse_2025}\nb{, where open-source activity remains concentrated in these regions. However, our sample slightly overrepresents experienced developers and contributors from Western regions, which may influence the generalizability of our findings.}

\end{tcolorbox}


\subsection{RQ1. Perception of Non-inclusive Terms by Demographics}
\label{sec:RQ1}

\nb{Throughout this section, statistical comparisons use the Mann-Whitney U test for two-group comparisons and the Kruskal-Wallis H test for three or more groups, both non-parametric methods appropriate for ordinal Likert data. Effect size is reported as the rank-biserial correlation $r$ (for U tests) or epsilon-squared $\varepsilon^2$ (for H tests), where $|r|$ or $\varepsilon^2 < .10$ = negligible, $.10$--$.29$ = small, $.30$--$.49$ = medium, and $\geq .50$ = large. All tests are two-sided ($\alpha = .05$).}

\begin{figure}[htbp]
\centering
\includegraphics[width=\columnwidth]{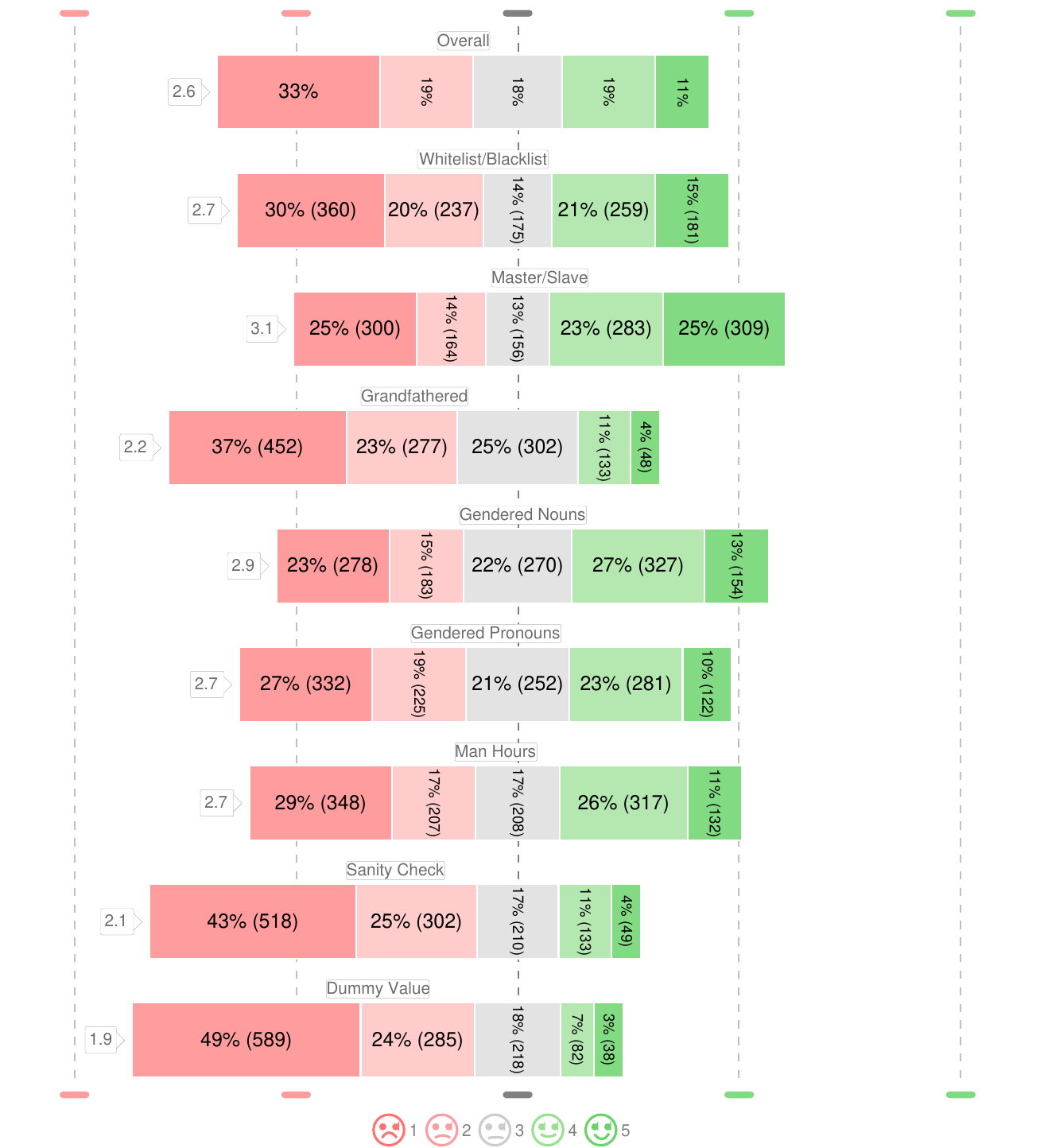}
\caption{\nb{Average Likert agreement scores with the statement ``I consider this term to be non-inclusive in software,'' across all respondents (1 = strongly disagree, 5 = strongly agree). The emoji scale represents the degree of agreement with this statement and does not indicate approval or endorsement of the terminology.}}
\label{fig:q1_non_inclusive_terms}
\end{figure}

The results below show the participants' average Likert agreement scores with the statement ``I consider this term to be non-inclusive in software'', which represents our hypothesis.

\textbf{Overall:} The average Likert scores across all participants indicate either neutrality or disagreement with our hypothesis that the studied terms have non-inclusive connotations in software (Figure~\ref{fig:q1_non_inclusive_terms}). Among our list of terms, ``master/slave'' had the highest average Likert score across all participants, slightly more than neutral, making them the terms that participants considered the most to be non-inclusive. ``Dummy value'' was perceived as the least non-inclusive term.

\begin{figure}[htbp]
\centering
\includegraphics[width=\columnwidth]{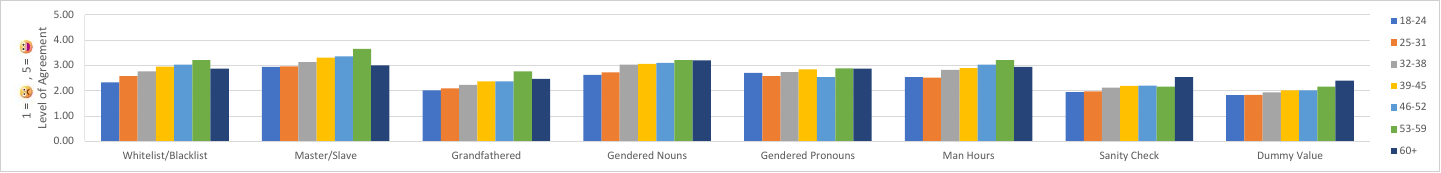}
\caption{\nb{The perception of non-inclusive terminology by age.}}
\label{fig:q1_age}
\end{figure}

\textbf{Age:} When we juxtapose the age distribution with the participants' opinions on these terms, we found that older participants are more receptive to the potential non-inclusive connotations of the terms (Figure~\ref{fig:q1_age}). Notably, participants in the age group of 53--59 tend to have a higher average agreement score with the hypothesis. \nb{This difference is statistically significant ($U = 124{,}346$, $p = .001$, $r = -.13$), with older respondents (39--60+) yielding a higher median score ($Mdn = 2.88$) than younger respondents (18--38, $Mdn = 2.50$), though the effect is small.}

\begin{figure}[htbp]
\centering
\includegraphics[width=\columnwidth]{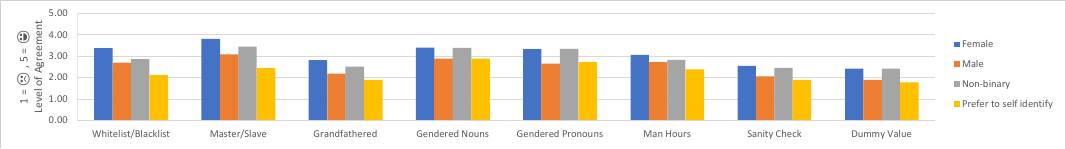}
\caption{The perception of non-inclusive terminology by gender.}
\label{fig:q1_gender}
\end{figure}

\textbf{Gender:} We also analyzed our dataset based on gender. Figure~\ref{fig:q1_gender} shows that there is a significant gender gap in the perception of the non-inclusive terms (10\%--27\%). Both women and non-binary participants tend to have a higher average agreement score with our hypothesis than other genders. \nb{A Kruskal-Wallis test confirmed a statistically significant difference across the three gender groups ($H(2) = 12.90$, $p = .002$, $\varepsilon^2 = .009$). Post-hoc pairwise comparisons revealed that the difference is driven primarily by female participants rating the terms as more non-inclusive than male participants ($U = 20{,}988$, $p < .001$, $r = -.29$; $Mdn$: Male $= 2.50$, Female $= 3.50$), a small but reliable effect. The difference between male and other/non-binary respondents did not reach significance ($p = .432$). At the per-term level, significant male--female differences were found for 7 of 8 terms (all $p < .05$), with the sole exception being ``Man Hours'' ($p = .100$).}

\begin{figure}[htbp]
\centering
\includegraphics[width=\columnwidth]{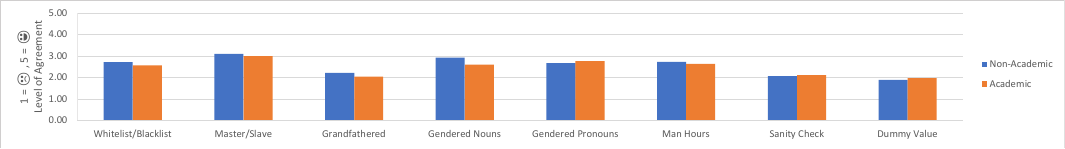}
\caption{The perception of non-inclusive terminology by academic and non-academic settings.}
\label{fig:q1_academic}
\end{figure}

\textbf{Occupation:} We compared non-academic and academic settings to see if there are disparities in the perception of our candidate non-inclusive terms (Figure~\ref{fig:q1_academic}). The differences are generally small across most terms. The most noticeable difference is in the use of gendered nouns, where \nb{participants in non-academic settings have higher average agreement that gendered nouns are non-inclusive than participants in academic settings ($Mdn = 3.0$ vs. $Mdn = 2.0$). A Mann-Whitney U test on the Gendered Nouns scores confirmed this difference is statistically significant ($U = 75{,}856$, $p = .009$, $r = .14$, small effect). For all other terms, no statistically significant difference was found between the two groups ($U = 62{,}696$, $p = .300$, $r = .06$ for the overall average), and the remaining descriptive differences should be interpreted with caution.}

\begin{figure}[htbp]
\centering
\includegraphics[width=\columnwidth]{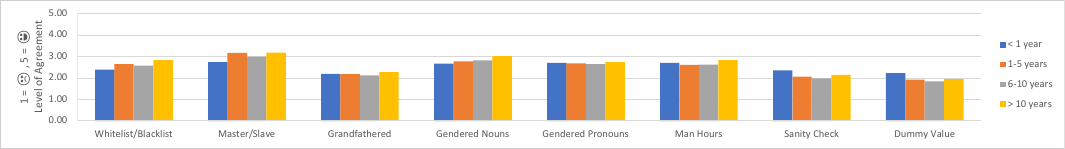}
\caption{The perception of non-inclusive terminology by programming experience.}
\label{fig:q1_experience}
\end{figure}

\textbf{Programming Experience:} The results presented in Figure~\ref{fig:q1_experience} indicate that the perception of most terms does not vary substantially across different levels of programming experience, with average agreement scores generally ranging between 2 and 3. However, participants with less than one year of programming experience have higher average agreement scores for ``Sanity Check'' and ``Dummy Value''. The number of participants in this group is relatively small (31 participants), representing only 11\% of the 18--24 age group, suggesting caution in interpretation. \nb{Consistent with this, a Mann-Whitney U test comparing low-experience (0--5 years) with high-experience (6+ years) respondents found no statistically significant difference in overall perception scores ($U = 107{,}934$, $p = .462$, $r = .03$).}

\begin{figure}[htbp]
\centering
\includegraphics[width=\columnwidth]{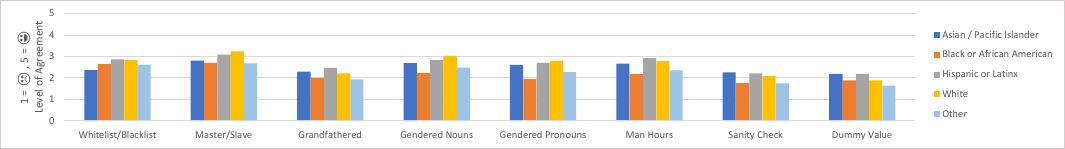}
\caption{The perception of non-inclusive terminology by ethnicity.}
\label{fig:q1_ethnicity}
\end{figure}

\textbf{Ethnicity:} Interestingly, White, Hispanic or Latinx, and Asian respondents tend to have a more neutral average score (Figure~\ref{fig:q1_ethnicity}), while other participants have a lower average agreement that these terms are non-inclusive. \nb{A Kruskal-Wallis test across four ethnicity groups did not yield a statistically significant overall difference ($H(3) = 4.48$, $p = .214$, $\varepsilon^2 = .001$). This result should be interpreted carefully: Black or African American respondents comprise only 17 participants, substantially limiting statistical power. Notably, the descriptive pattern shows a lower median for Black or African American respondents ($Mdn = 1.38$) relative to White ($Mdn = 2.62$), Asian/Pacific Islander ($Mdn = 2.62$), and Hispanic or Latinx respondents ($Mdn = 2.75$), which may reflect survivorship bias, as discussed in Section~\ref{sec:discussion}}.

\begin{figure}[htbp]
\centering
\includegraphics[width=\columnwidth]{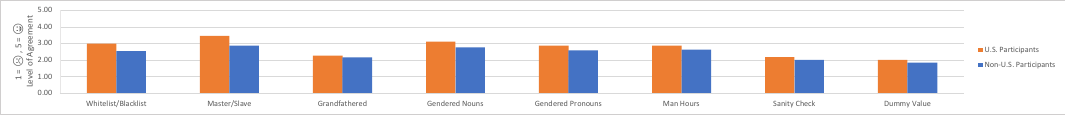}
\caption{The perception of non-inclusive terminology by U.S. and non-U.S. participants.}
\label{fig:q1_us_non_us}
\end{figure}

\textbf{U.S. vs. non-U.S. participants:} We found that U.S. participants, who are mostly native English speakers (90\%), demonstrate higher sensitivity to non-inclusive language than non-U.S. participants (only 19\% are native English speakers), as shown in Figure~\ref{fig:q1_us_non_us}. We also observed a similar trend when comparing native and non-native English speakers. These results highlight the influence of cultural and linguistic differences on perceptions of non-inclusive language in software engineering.  \nb{Both differences are statistically significant. U.S. participants scored higher than non-U.S. participants ($U = 205{,}492$, $p < .001$, $r = .16$; $Mdn$: U.S.$= 2.75$, Non-U.S. $= 2.38$), and native English speakers scored higher than non-native speakers ($U = 218{,}036$, $p < .001$, $r = .19$; $Mdn$: Native $= 2.88$, Non-native $= 2.38$), with small effects in both cases. At the per-term level, significant U.S./non-U.S. differences were found for 7 of 8 terms; the sole exception was ``Grandfathered'' ($p = .135$), suggesting its specific historical connotations may be less widely recognized even within the U.S. context.}

\begin{figure}[htbp]
\centering
\includegraphics[width=\columnwidth]{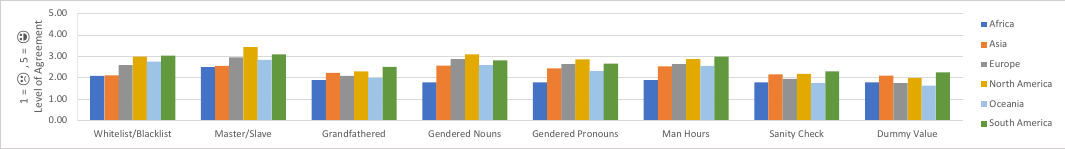}
\caption{The perception of non-inclusive terminology by continent residency.}
\label{fig:q1_residency}
\end{figure}

\textbf{Residency:} By residency, participants from North America and South America are more receptive to the non-inclusive connotations of the terms (Figure~\ref{fig:q1_residency}).

\begin{tcolorbox}[mpd, title={Summary (RQ1)}]
This research investigated how developers from different demographic groups perceive non-inclusive terminology in software. Overall, most participants were neutral or disagreed with the hypothesis that the studied terms were non-inclusive. Among the terms examined, ``master/slave'' was considered the most non-inclusive, while ``dummy value'' was perceived as the least.

\nb{Statistically significant differences in perception were confirmed for gender ($H(2) = 12.90$, $p = .002$), age ($U$, $p = .001$), U.S./non-U.S. residence ($U$, $p < .001$), and native language ($U$, $p < .001$), though all effect sizes were small ($r$ or $\varepsilon^2 < .30$).} Older participants showed greater awareness of potential non-inclusive connotations, and women reported higher average agreement with the hypothesis than male participants. 

\nb{By contrast, occupation ($p = .300$) and programming experience ($p = .462$) did not yield statistically significant differences, and the ethnicity comparison was underpowered due to the small number of Black or African American respondents ($n = 17$).} 

U.S.-based and native English-speaking respondents were more sensitive to non-inclusive language than their non-U.S. and non-native English-speaking counterparts, highlighting the role of cultural and linguistic context in shaping perceptions of non-inclusive terminology in software engineering.

\nb{Under Holm--Bonferroni correction ($k = 11$ family), the significant gender (both the overall test and the Male vs.\ Female comparison), age, U.S./non-U.S.\ residence, and native-language differences all survive (all $p \leq .002$), confirming their robustness to multiple-comparison adjustment. The only originally significant result that does not survive is the academic vs.\ non-academic difference on Gendered Nouns ($p = .009$), which falls just above its corrected threshold (.008) and should therefore be interpreted with caution.}
\end{tcolorbox}

\subsection{RQ2a. The Impact of Non-inclusive Terms on Participants who Disagreed with our Hypothesis}
\label{sec:RQ2a}

To address RQ2, we divided the participants into two categories based on the number of instances where they disagreed, agreed, or remained neutral with our hypothesis. In this section (RQ2a), we focused on participants with more instances of terms for which they disagreed or remained neutral with our hypothesis. This allowed us to examine the impact of non-inclusive terms on participants who may be less sensitive to these issues or hold different perspectives on language use in software. In Section~\ref{sec:RQ2b}, we focused on participants with more instances of terms for which they agreed with our hypothesis, allowing us to explore the relationship between sensitivity to non-inclusive language and perspectives towards inclusive language use in software.

\nb{Throughout this section, comparisons between two groups on categorical reason-selection variables use Pearson's chi-square test ($\chi^2$) or Fisher's exact test where expected cell counts fall below 5. Effect size is reported as Cram\'{e}r's $V$, where $V < .10$ = negligible, $.10$--$.29$ = small, $.30$--$.49$ = medium, and $\geq .50$ = large. All tests are two-sided ($\alpha = .05$).}

\begin{figure}[htbp]
\centering
\includegraphics[width=0.5\linewidth]{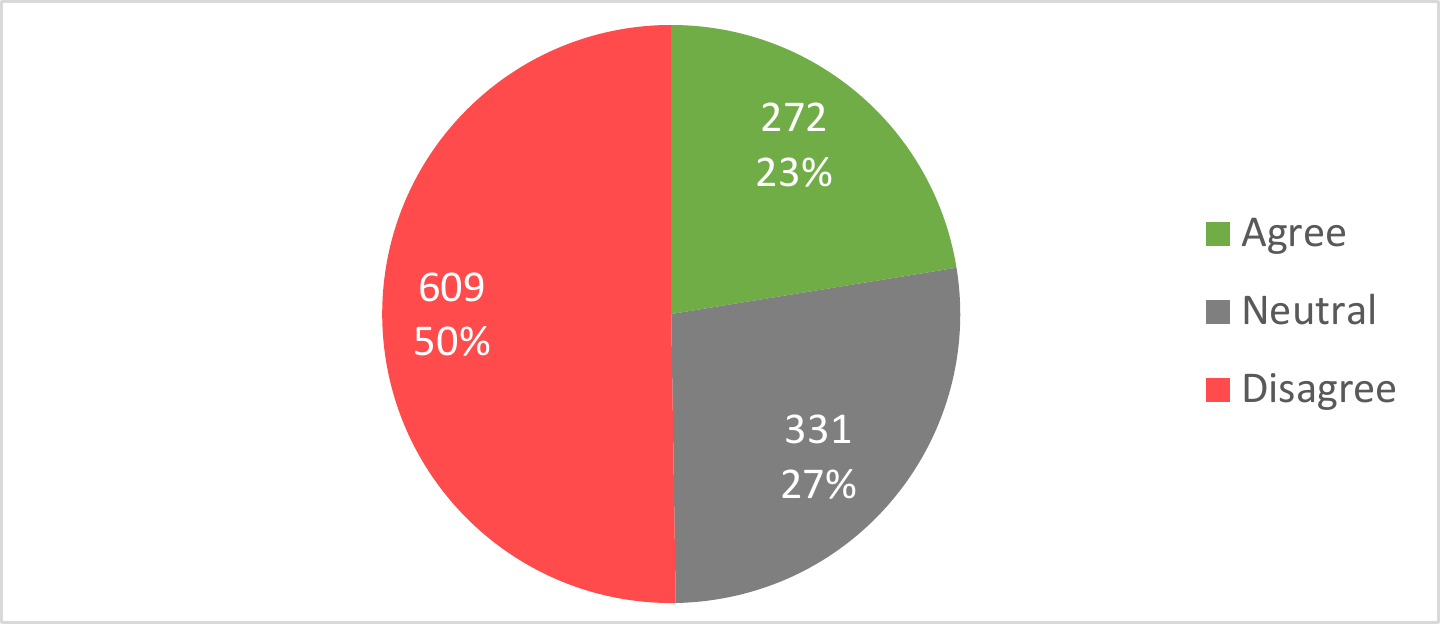}
\caption{Distribution of participants based on the majority of instances where they disagreed, agreed or were neutral with our hypothesis.}
\label{fig:q2a}
\end{figure}

Specifically, in RQ2a we focused on the 940 participants (77\% of the total sample) who disagreed or remained neutral with our hypothesis for more terms than they agreed. Figure~\ref{fig:q2a} depicts the distribution of participants based on the majority of instances where they disagreed, agreed, or remained neutral with our hypothesis. By examining this group we aimed to gain a better understanding of the potential impact of non-inclusive terms on participants who may not prioritize or be sensitive to issues of inclusivity in software development.

\begin{figure}[htbp]
\centering
\includegraphics[width=\columnwidth]{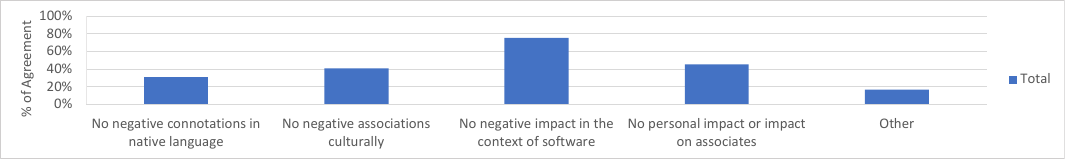}
\caption{Reasons why participants disagree that the candidate terms are non-inclusive.}
\label{fig:q2a_general}
\end{figure}

\textbf{Overall:} Figure~\ref{fig:q2a_general} shows the responses from the participants who disagreed or remained neutral with our hypothesis for more terms than they agreed (\eg they believe more terms in our list do NOT have non-inclusive connotations in software or they were neutral). The results overall show that most participants do not believe these terms have a negative impact in the context of software (76\%). Also, almost half of them (46\%) see that these terms do not personally impact them or their associates.

\begin{figure}[htbp]
\centering
\includegraphics[width=\columnwidth]{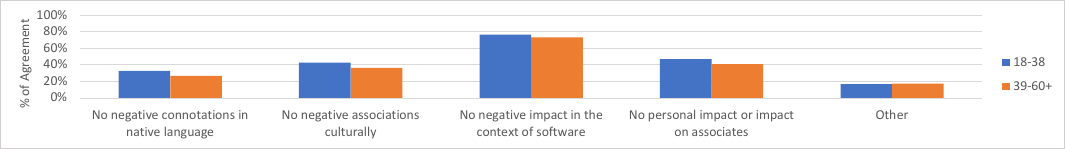}
\caption{Reasons why participants with different age groups disagreed or remained neutral that most candidate terms are non-inclusive.}
\label{fig:q2a_age}
\end{figure}

\textbf{Age:} The data presented in Figure~\ref{fig:q2a_age} implies that participants between the ages of 18--38 are less likely to be impacted by non-inclusive terms than older participants. Specifically, younger participants tend to be less affected by the non-inclusive terms. \nb{Chi-square tests on the four reason categories found no statistically significant differences between age groups on any of the four reason categories (all $p > .08$, all $V < .06$). These descriptive differences should therefore be interpreted with caution.}

\begin{figure}[htbp]
\centering
\includegraphics[width=\columnwidth]{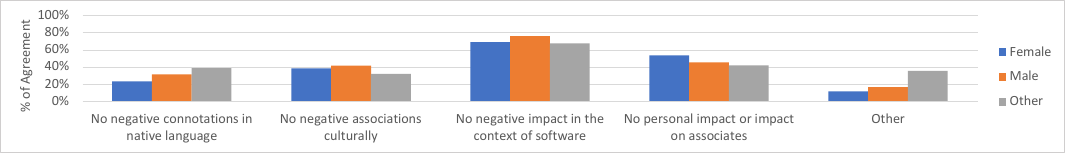}
\caption{Reasons why participants with different genders disagreed or remained neutral that most candidate terms are non-inclusive.}
\label{fig:q2a_gender}
\end{figure}

\textbf{Gender:} Based on the data presented in Figure~\ref{fig:q2a_gender}, non-inclusive terms have a lower likelihood of causing personal impact or impacting other associates for female participants who disagree or remain neutral with the hypothesis than their male or other counterparts. Specifically, the difference in likelihood is approximately 8 percentage points and 12 percentage points, respectively. \nb{However, chi-square tests across all four reason categories found no statistically significant differences between male and female participants (all $p > .48$, all $V < .03$) or between male and other-gender participants (all $p > .37$, all $V < .03$), likely due to the small number of female ($n = 26$) and other-gender ($n = 31$) disagreers in this subgroup. These descriptive differences should therefore be interpreted with caution.}

\begin{figure}[htbp]
\centering
\includegraphics[width=\columnwidth]{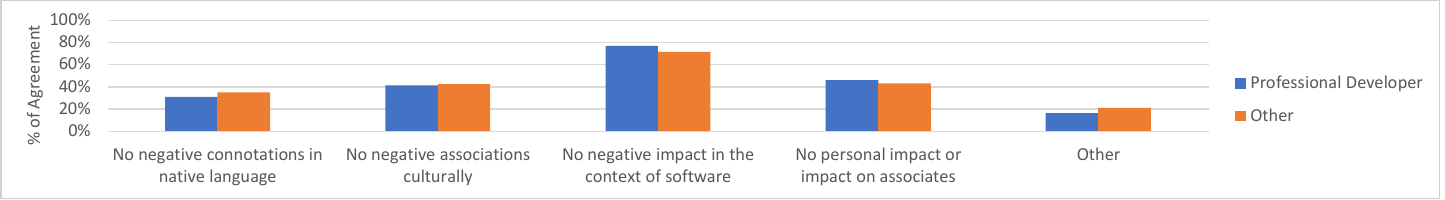}
\caption{Reasons why participants with different primary occupations disagreed or remained neutral that most candidate terms are non-inclusive.}
\label{fig:q2a_occupation}
\end{figure}

\textbf{Occupation:} Based on the data presented in Figure~\ref{fig:q2a_occupation}, non-inclusive terms tend to have a lower likelihood of causing a negative impact in the context of software for professional developer participants who disagree or remain neutral with the hypothesis than other participants, including participants in academia. \nb{A chi-square test found a statistically significant difference between academic and non-academic participants on the ``no negative connotation in native language'' reason ($\chi^2(1) = 4.55$, $p = .033$, $V = .070$), with academic participants more likely to endorse this reason (41.4\%) than non-academic participants (30.3\%), though the effect size is negligible. No statistically significant differences were found for the remaining three reason categories (all $p > .44$, all $V < .03$). The descriptive patterns for the other reasons should be interpreted as tendencies rather than reliable differences.}

\begin{figure}[htbp]
\centering
\includegraphics[width=\columnwidth]{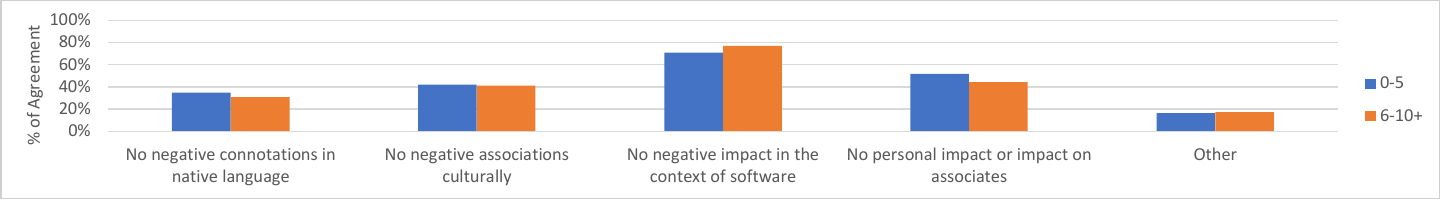}
\caption{Reasons why participants with different programming experience groups disagreed or remained neutral that most candidate terms are non-inclusive.}
\label{fig:q2a_experience}
\end{figure}

\textbf{Programming experience:} The information depicted in Figure~\ref{fig:q2a_experience} shows that non-inclusive terms tend to have a lower likelihood of causing a negative impact in the context of software for highly experienced participants who disagree or remain neutral with the hypothesis than participants with low programming experience. \nb{Chi-square tests found no statistically significant differences between low-experience (0--5 years) and high-experience (6+ years) participants on any reason category (all $p > .08$, all $V < .06$), consistent with the non-significant experience result in RQ1 ($p = .462$).}

\begin{figure}[htbp]
\centering
\includegraphics[width=\columnwidth]{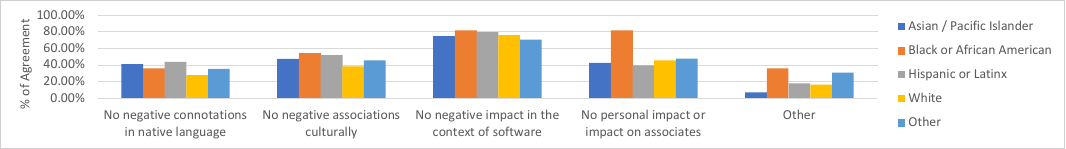}
\caption{Reasons why participants with different ethnicities disagreed or remained neutral that most candidate terms are non-inclusive.}
\label{fig:q2a_ethnicity}
\end{figure}

\textbf{Ethnicity:} Non-inclusive terms have a significantly lower likelihood of causing personal impact or impacting other associates on Black or African American participants who disagree or remain neutral with the hypothesis than other ethnicities (Figure~\ref{fig:q2a_ethnicity}). For instance, the difference in impact ranges from 34 to 42 percentage points higher than for Hispanic/Latinx, Asian/Pacific Islander, White, and other races participants. \nb{However, this comparison must be interpreted with caution: only 11 Black or African American respondents fell within the RQ2a subgroup ($n = 11$), rendering formal chi-square tests unreliable for this cell. A Fisher's exact test on the ``no personal impact'' reason comparing Black or African American respondents with all other ethnicities combined reached statistical significance ($p = .034$, $V = .069$), though the effect size is negligible. The very small sample substantially limits statistical power, and this finding warrants replication with larger, more representative samples.}

\begin{figure}[htbp]
\centering
\includegraphics[width=\columnwidth]{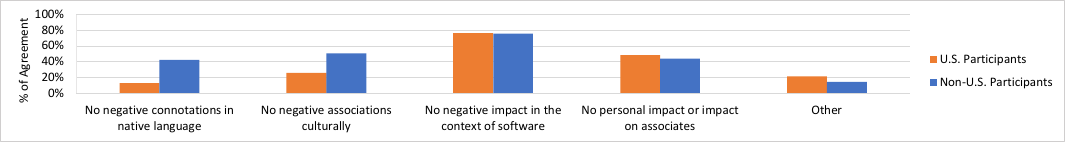}
\caption{Reasons why U.S. and non-U.S. participants disagreed or remained neutral that most candidate terms are non-inclusive.}
\label{fig:q2a_us_non_us}
\end{figure}

\textbf{U.S. vs. non-U.S. participants:} Figure~\ref{fig:q2a_us_non_us} shows why U.S. and non-U.S. participants disagree or remain neutral with our hypothesis. We noticed that non-U.S. participants are 24 percentage points more likely than U.S. participants to report no negative cultural associations with these candidate non-inclusive terms. They are also 29 percentage points more likely than U.S. participants to report no negative connotations with these candidate non-inclusive terms because their native language carries no such associations. These findings suggest that certain non-inclusive terms may be specific to the U.S. context and may not apply to other cultural contexts. \nb{Both differences are statistically significant. The 29 percentage point difference in native-language connotations ($\chi^2(1) = 84.16$, $p < .001$, $V = .299$) and the 24 percentage point difference in cultural associations ($\chi^2(1) = 51.87$, $p < .001$, $V = .235$) each represent small effects, confirming that these are not sampling artefacts but reliable cross-cultural differences. These results converge with the significant U.S./non-U.S. difference found in RQ1 ($U = 205{,}492$, $p < .001$, $r = .16$).}

\begin{figure}[htbp]
\centering
\includegraphics[width=\columnwidth]{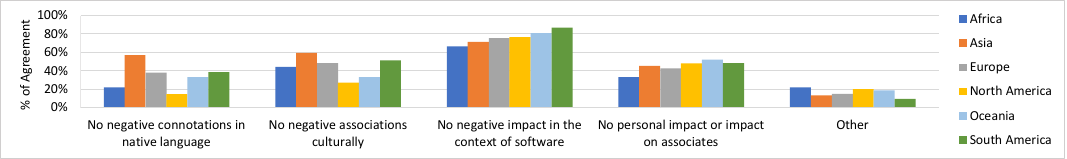}
\caption{Reasons why participants with different continent residencies disagreed or remained neutral that most candidate terms are non-inclusive.}
\label{fig:q2a_residency}
\end{figure}

\textbf{Residency:} Analyzing the results by continent of residency, Figure~\ref{fig:q2a_residency} shows that nearly all participants from South America felt no negative impact of non-inclusive terms in the context of software. Participants from Asia, Europe, and South America are more likely \nb{than those from North America} to have no negative cultural associations with these candidate non-inclusive terms. Respondents from Asia, Europe, South America, Oceania, and Africa are more likely \nb{than North American respondents} to have no negative connotations of these candidate non-inclusive terms due to their native languages.

\textbf{Other Reasons:} Two authors utilized an open coding method to categorize additional reasons provided by participants in response to the optional open-ended question about their disagreement or neutrality regarding non-inclusive terminology. After resolving any discrepancies, the authors identified ten distinct categories for these reasons, presented in \hyperref[tab:other_reasons_q11]{Table}~\ref{tab:other_reasons_q11}. Among the participants who provided other reasons, there was a consensus of about 38\% on two specific explanations: the candidate terms can be harmful depending on the context in which they are used, and developers do not intend to use these terms offensively. The next most commonly agreed-upon reason (22\%) was that participants were not aware of or did not consider the discriminatory use of some terms.

\begin{table}
  \centering
  \caption{The additional reasons provided by participants in response to the optional open-ended question about the disagreement or remaining neutral about not considering the candidate terms as non-inclusive or discriminatory}
  \label{tab:other_reasons_q11}
  \resizebox{\columnwidth}{!}{
    \begin{tabular}{lcc}
      \toprule
      \textbf{Category of Other Reasons} & \textbf{Count} & \% \textbf{of agreement} \\
      \midrule
      Not used in an offensive way                      & 82 & 38\% \\
      Depends on context                                & 81 & 38\% \\
      Not aware of the discriminatory nature            & 48 & 22\% \\
      People read too much into negative meanings       & 40 & 19\% \\
      Some of these terms are offensive                 & 27 & 13\% \\
      Legacy language                                   & 16 & 7\%  \\
      There are other issues to combat discrimination   & 10 & 5\%  \\
      Never heard of some words                         &  7 & 3\%  \\
      Others may find the terms offensive               &  6 & 3\%  \\
      Language-dependent                                &  6 & 3\%  \\
      \bottomrule
    \end{tabular}
  }
\end{table}

\begin{tcolorbox}[mpd, title={Summary (RQ2a)}] This section examines the perceived impact of non-inclusive terminology among participants who disagreed or remained neutral regarding the hypothesis, providing insight into how developers who may not prioritize inclusivity issues interpret the potential effects of such language. Overall, most participants in this category (76\%) did not believe that non-inclusive terminology has a negative impact in the context of software development, and nearly half (46\%) reported that these terms do not personally affect them or their associates.

\nb{Statistically significant demographic differences were found primarily for the U.S./non-U.S. dimension, with non-U.S. participants substantially more likely to cite native-language ($\chi^2(1) = 84.16$, $p < .001$, $V = .299$) and cultural ($\chi^2(1) = 51.87$, $p < .001$, $V = .235$) reasons for their neutrality, both small effects. A significant but negligible-effect difference was also found for occupation on the ``no negative connotation in native language'' reason ($\chi^2(1) = 4.55$, $p = .033$, $V = .070$), with academic participants more likely to endorse this reason than non-academic participants. A significant Fisher's exact result was found for Black or African American respondents on ``no personal impact'' ($p = .034$, $V = .069$), though this must be interpreted with extreme caution given the very small subgroup size ($n = 11$). Differences by age, gender, and programming experience on the reason categories were not statistically significant (all $p > .05$), though the descriptive pattern of younger participants and professional developers being less likely to perceive impact is consistent with the broader RQ1 findings.}

Participants who disagreed or remained neutral offered several explanations for their views, including that the meaning of these terms depends on context, that developers typically do not use them with harmful intent, and that some were unaware of the potentially discriminatory origins or implications of certain terms.

\nb{Under Holm--Bonferroni correction ($k = 28$ family), only the two U.S./non-U.S.\ differences survive (both $p < .001$). The occupation ($p = .033$) and ethnicity ($p = .034$) findings do not meet the corrected threshold and should be interpreted with additional caution.}
\end{tcolorbox}


\subsection{RQ2b. The Impact of Non-inclusive Terms on Participants who Agreed with our Hypothesis}
\label{sec:RQ2b}

In this section, we examined the responses of participants who agreed with our hypothesis for a majority of the terms. Specifically, we focused on the 272 participants (23\% of the total sample) who agreed with our hypothesis for more terms than they disagreed or remained neutral (Figure~\ref{fig:q2a}). By examining this group we aimed to gain a better understanding of the potential impact of non-inclusive terms on participants who may prioritize or be sensitive to issues of inclusivity in software development.

\nb{Throughout this section, differences in per-dimension impact rates between demographic groups are tested with Pearson's chi-square ($\chi^2$) or Fisher's exact test where expected cell counts fall below 5. Effect size is reported as Cram\'{e}r's $V$. Overall impact across all five dimensions is also compared using the Mann-Whitney U test (MWU) on the average impact score, with effect size reported as rank-biserial correlation $r$. Thresholds: $V$ or $|r| < .10$ = negligible, $.10$--$.29$ = small, $.30$--$.49$ = medium, $\geq .50$ = large. All tests are two-sided ($\alpha = .05$).}

\begin{figure}[htbp]
\centering
\includegraphics[width=\columnwidth]{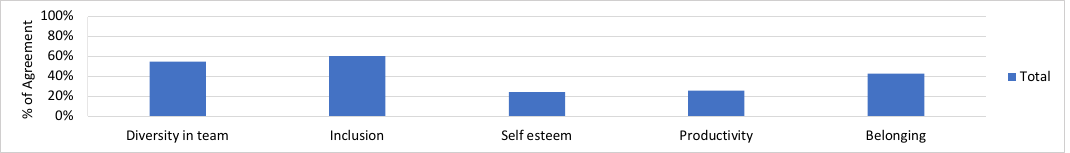}
\caption{The aspects that participants considered the non-inclusive terms to impact in their lives.}
\label{fig:q2b_general}
\end{figure}

\textbf{Overall:} Figure~\ref{fig:q2b_general} illustrates the aspects these participants attributed to the impact of non-inclusive terms on their lives. Over 50\% of the participants in this group expressed that non-inclusive terms negatively impact their sense of inclusion and diversity within their team. Furthermore, over 40\% of participants felt that non-inclusive terms negatively affect their sense of belonging, while approximately 30\% perceived potential negative impacts on their productivity and self-esteem.

\begin{figure}[htbp]
\centering
\includegraphics[width=\columnwidth]{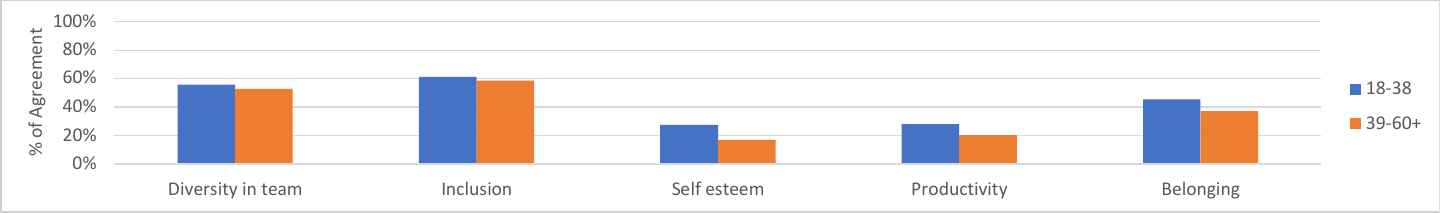}
\caption{The aspects that participants with different age groups considered the non-inclusive terms to impact in their lives.}
\label{fig:q2b_age}
\end{figure}

\textbf{Age:} Our analysis indicates that younger participants (aged 18--38) who agreed with our hypothesis were more impacted by non-inclusive terms than older participants who agreed with our hypothesis. Specifically, younger respondents reported higher negative impacts on their sense of belonging (by 8 percentage points), productivity (by 8 percentage points), and self-esteem (by 10 percentage points) than older participants, as shown in Figure~\ref{fig:q2b_age}. \nb{A Mann-Whitney U test on the average impact score confirmed a statistically significant overall difference ($U = 9{,}512$, $p = .024$, $r = .17$; $Mdn$: 18--38 $= 3.20$, 39--60+ $= 3.00$), with a small effect. However, chi-square tests on individual dimensions found that none of the per-dimension differences reached significance (all $p > .08$, all $V < .11$), suggesting the effect is distributed broadly across dimensions rather than concentrated in any single one.}

\begin{figure}[htbp]
\centering
\includegraphics[width=\columnwidth]{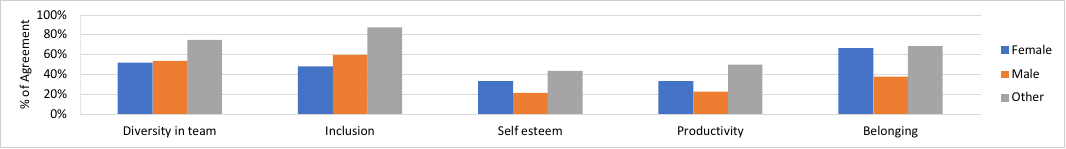}
\caption{The aspects that participants with different genders considered the non-inclusive terms to impact in their lives.}
\label{fig:q2b_gender}
\end{figure}

\textbf{Gender:} We found that female and other gender respondents who agreed with our hypothesis reported a negative impact on their sense of belonging (by 29 to 31 percentage points), self-esteem (by 12 to 22 percentage points), and productivity (by 11 to 27 percentage points) than male respondents (Figure~\ref{fig:q2b_gender}). Additionally, other gender participants, including non-binary, reported higher negative impacts on their sense of inclusion and diversity in teams, with differences of 39 percentage points and 23 percentage points, respectively, relative to female participants. \nb{A Kruskal-Wallis test on the average impact score confirmed a statistically significant overall gender difference ($H(2) = 7.37$, $p = .025$, $\varepsilon^2 = .020$). Pairwise Mann-Whitney U tests showed this is driven by the Male vs. Other/Non-binary comparison ($U = 1{,}184$, $p = .018$, $r = -.353$, medium effect; $Mdn$: Male $= 3.00$, Other $= 3.80$), while the Male vs. Female difference did not reach significance on the average impact score ($U = 2{,}536$, $p = .126$, $r = -.180$; $Mdn$: Male $= 3.00$, Female $= 3.20$). Per-dimension chi-square tests revealed that the significant differences are concentrated in specific dimensions. For belonging, both female ($\chi^2(1) = 7.07$, $p = .008$, $V = .17$; Male = 38\% vs. Female = 67\%) and other-gender participants ($\chi^2(1) = 4.68$, $p = .031$, $V = .14$; Male = 38\% vs. Other = 69\%) reported significantly higher impact than male respondents, with small effects. For productivity, other-gender participants also reported significantly higher impact than male respondents (Fisher $p = .030$, $V = .14$; Male = 23\% vs. Other = 50\%). The female--other-gender difference on inclusion was significant with a medium effect ($\chi^2(1) = 5.08$, $p = .024$, $V = .34$; Female = 48\% vs. Other = 88\%). By contrast, per-dimension differences on self-esteem and productivity between male and female respondents did not reach significance ($p = .247$ and $p = .324$, respectively), nor did the female--other-gender difference on diversity ($p = .239$). These findings indicate that the gender effect is strongest for sense of belonging and, among other-gender participants, for productivity and inclusion.}

\begin{figure}[htbp]
\centering
\includegraphics[width=\columnwidth]{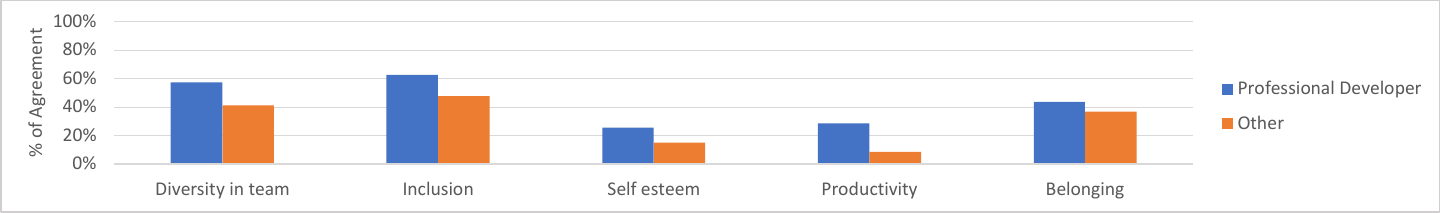}
\caption{The aspects that participants with different primary occupations considered the non-inclusive terms to impact in their lives.}
\label{fig:q2b_occupation}
\end{figure}

\textbf{Occupation:} Our analysis points out that professional developer participants who agreed with our hypothesis were more impacted by non-inclusive terms than participants with other primary occupations, including academia. Specifically, professional developer respondents reported higher negative impacts on their sense of productivity (by 20 percentage points), diversity in teams (by 16 percentage points), inclusion (by 15 percentage points), self-esteem (by 10 percentage points), and belonging (by 7 percentage points) than other participants, as illustrated in Figure~\ref{fig:q2b_occupation}. \nb{However, a Mann-Whitney U test on the average impact score found no statistically significant difference between professional developers and academic participants ($U = 3{,}330$, $p = .195$, $r = .16$). Chi-square tests on individual dimensions likewise found no significant differences (all $p > .09$, all $V < .11$). The descriptive differences should therefore be interpreted with caution, particularly given the small number of academic agreers ($n = 23$).}

\begin{figure}[htbp]
\centering
\includegraphics[width=\columnwidth]{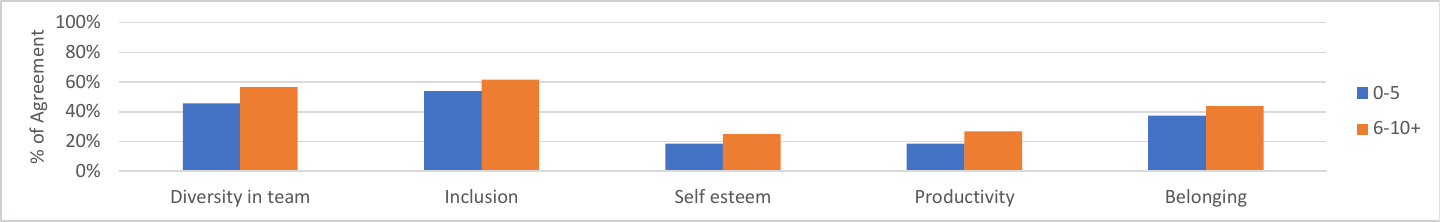}
\caption{The aspects that participants with different programming experience groups considered the non-inclusive terms to impact in their lives.}
\label{fig:q2b_experience}
\end{figure}

\textbf{Programming experience:} Our analysis indicates that highly experienced participants who agreed with our hypothesis reported higher negative impacts on their sense of diversity in teams (by 11 percentage points), productivity (by 8 percentage points), inclusion (by 7 percentage points), self-esteem (by 6 percentage points), and belonging (by 6 percentage points) than low-experience participants, as illustrated in Figure~\ref{fig:q2b_experience}. \nb{A Mann-Whitney U test on the average impact score found no statistically significant difference between low-experience (0--5 years) and high-experience (6+ years) participants ($U = 5{,}900$, $p = .289$, $r = -.10$). Chi-square tests on individual dimensions were likewise non-significant (all $p > .23$, all $V < .08$). The observed descriptive tendencies should be treated with caution.}

\begin{figure}[htbp]
\centering
\includegraphics[width=\columnwidth]{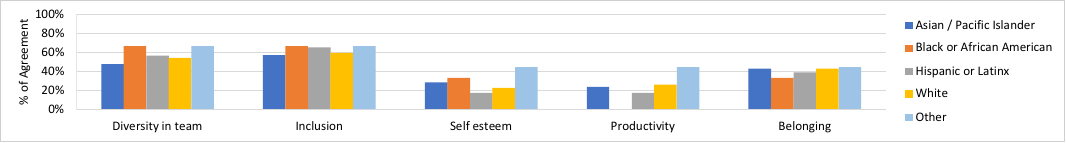}
\caption{The aspects that participants with different ethnicities considered the non-inclusive terms to impact in their lives.}
\label{fig:q2b_ethnicity}
\end{figure}

\textbf{Ethnicity:} Figure~\ref{fig:q2b_ethnicity} shows that Black or African American and other race respondents who agreed with our hypothesis felt that non-inclusive terms could negatively affect team diversity (10--19 percentage points more than other groups). Likewise, non-inclusive terms tended to negatively impact their self-esteem more than other groups by 5--17 percentage points and 16--27 percentage points, respectively. Also, non-inclusive terms tended to have higher negative impacts on productivity for other races than other groups by more than 18 percentage points. Note that we did not have data points for productivity from Black or African American respondents. \nb{However, all per-dimension comparisons between Black or African American respondents and other ethnicities were non-significant (all Fisher $p > .34$), and must be interpreted with considerable caution given the very small number of Black or African American agreers in this subgroup ($n = 6$). This finding warrants replication with a larger, more representative sample.}

\begin{figure}[htbp]
\centering
\includegraphics[width=\columnwidth]{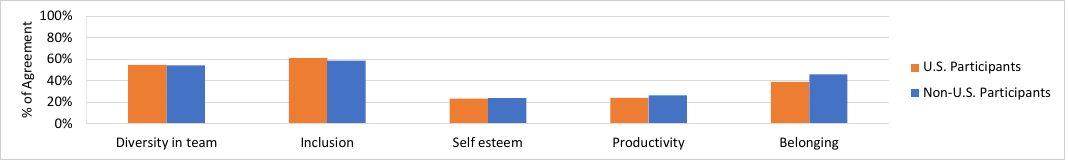}
\caption{The aspects that U.S. and non-U.S. participants considered the non-inclusive terms to impact in their lives.}
\label{fig:q2b_us_non_us}
\end{figure}

\textbf{U.S. vs. non-U.S. participants:} Figure~\ref{fig:q2b_us_non_us} shows that non-U.S. respondents who agreed with the hypothesis felt a more negative impact of non-inclusive terms on their sense of belonging (by 7 percentage points). \nb{A Mann-Whitney U test on the average impact score found no statistically significant difference between U.S. and non-U.S. participants ($U = 9{,}409$, $p = .794$, $r = .02$; $Mdn$: U.S. $= 3.20$, Non-U.S. $= 3.20$). The 7 percentage point belonging difference was also non-significant ($\chi^2(1) = 1.06$, $p = .302$, $V = .06$). Among participants who recognized the non-inclusive nature of the terms, the level of reported negative impact was equivalent regardless of cultural background.}

\begin{figure}[htbp]
\centering
\includegraphics[width=\columnwidth]{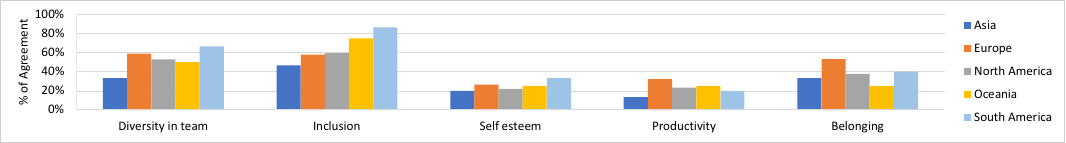}
\caption{The aspects that participants with different continent residencies considered the non-inclusive terms to impact in their lives.}
\label{fig:q2b_residency}
\end{figure}

\textbf{Residency:} According to Figure~\ref{fig:q2b_residency}, participants from South America who agreed with our hypothesis reported the most significant negative impact of non-inclusive language on their sense of inclusion (12--40 percentage points), diversity in teams (7--33 percentage points), and self-esteem (7--13 percentage points). On the other hand, European participants who agreed with our hypothesis reported feeling the most negative impact on their sense of belonging (13--28 percentage points) and productivity (8--19 percentage points). Notably, we excluded African participants who agreed with our hypothesis from our analysis due to the small sample size (only one participant), and thus, we cannot draw meaningful conclusions based on this group.

\begin{tcolorbox}[mpd, title={Summary (RQ2b)}]
This section examines the perceived impact of non-inclusive terminology among participants who agreed with the hypothesis, providing insight into how developers who are more attuned to inclusivity issues experience such language. More than half of these participants reported that non-inclusive terms negatively affect their sense of inclusion and team diversity. Over 40\% indicated adverse effects on their sense of belonging, and roughly 30\% perceived negative impacts on productivity and self-esteem.

\nb{Statistically significant differences were confirmed for gender and age. Younger respondents reported stronger overall negative impact than older participants ($U = 9{,}512$, $p = .024$, $r = .17$), though no individual dimension reached significance on its own. For gender, the overall Kruskal-Wallis test was significant ($H(2) = 7.37$, $p = .025$, $\varepsilon^2 = .020$), with the key per-dimension findings being: female respondents reported significantly higher impact on belonging than males ($\chi^2(1) = 7.07$, $p = .008$, $V = .17$); other-gender respondents reported significantly higher impact on belonging ($p = .031$, $V = .14$) and productivity ($p = .030$, $V = .14$) than males; and other-gender respondents reported significantly higher impact on inclusion than female respondents ($p = .024$, $V = .34$, medium effect). Gender differences on self-esteem and productivity between male and female respondents were descriptive only and did not reach significance.}

\nb{By contrast, occupation ($p = .195$), programming experience ($p = .289$), and U.S./non-U.S. residence ($p = .794$) did not yield statistically significant differences on overall impact, and per-dimension tests were likewise non-significant for all three. Ethnicity comparisons were non-significant and underpowered due to the very small number of Black or African American agreers ($n = 6$).} Racial and regional descriptive differences are reported but should be treated with caution pending replication in larger, more representative samples.

\nb{Under Holm--Bonferroni correction ($k = 9$ for overall-average tests; $k = 40$ for per-dimension tests), none of the RQ2b findings survive. These results are therefore exploratory, consistent with the small effect sizes and limited subgroup sizes noted throughout.}
\end{tcolorbox}

\section{Discussion}\label{sec:discussion}

The results from our survey of 1,212 software developers provide an empirical foundation for many of the anecdotal opinions that have circulated online regarding non-inclusive terminology in software artifacts. The findings highlight several overlapping themes concerning awareness, perceived priorities, gender-related experiences, and the importance of contextual interpretation.

\textbf{Awareness and cultural context.}  
Many non-U.S. developers and non-native English speakers appeared to have limited awareness of the historical or cultural connotations associated with certain non-inclusive terms. This finding aligns with the broader observation that the cultural origins of such terminology are often specific to the United States. 

\nb{A possible explanation for this difference is that U.S.-based participants may have greater exposure to ongoing public and professional discourse around inclusive language, diversity, and social responsibility. Over the past decade, such topics have received significant attention in U.S.-based industry initiatives, media, and workplace policies, which may influence how developers interpret terminology and its potential impact. In contrast, developers in other regions may have had less exposure to these discussions or may interpret such terminology primarily through a technical rather than socio-historical lens.}

For example, Petr Baudis, who selected the term ``master'' for the main Git reference in 2005, later explained that his intention was to evoke ``master recording,'' not a master–slave relationship \citep{landau_2020}. Beyond the U.S. context, terms like ``slave'' are often associated with punishment for crimes or prisoners of war rather than with racialized histories, making it difficult for some non-U.S. developers to grasp the term's connection to racial oppression. 

\nb{It should be noted, however, that the legacy of slavery and racial oppression is not exclusively a U.S.\ phenomenon: many European nations with colonial histories---including the United Kingdom, France, Spain, Portugal, and the Netherlands---also engaged in large-scale enslavement, and structural racism rooted in this history persists in those societies today. The cultural salience of such terminology may therefore vary considerably even within non-U.S.\ contexts.}  

Other expressions, such as ``grandfathered,'' are even more deeply rooted in U.S. history and legislation, and their connotations may be opaque to international audiences. Illustrative participant comments include: ``\textit{I don't know the meaning and connotations of the term},'' and ``\textit{Black is not associated with race in Japan}.'' Several respondents also noted that it may be inappropriate to impose U.S.-specific standards of political correctness on societies with different cultural norms or linguistic traditions. Such remarks highlight the challenge of developing globally inclusive language standards in a field as international as software engineering.

\textbf{Perceived priorities and misdirected efforts.}  
A subset of participants acknowledged the \nb{potential} offensiveness of certain terms but argued that efforts to improve diversity, equity, and inclusion (DEI) should focus on structural inequities (such as pay gaps and promotion barriers) rather than on terminology alone. As one respondent observed, ``\textit{I am aware of the issues of discrimination, but I'm kinda old and very desensitized to these issues. I have friends who are strong DEI advocates who have brought many of these issues to light and fully agree that we as a society (not just as software engineers) need to pay attention to our use of language and its impact on our community.}''  

Others viewed terminology change as a low-cost but symbolically meaningful action that can accompany deeper reforms. One participant stated, ``\textit{I believe we need to focus on addressing underlying issues of empowerment and fairness rather than focusing on changing our language without changing our system.}'' Another countered that ``\textit{It is important to address these in software. I'm aware that my associates or I may not have been affected, but others might be. To accommodate them, the industry can start to switch over to other terms.}''  

At the same time, several participants highlighted the technical and logistical challenges associated with changing long-established terminology. As one developer pointed out, ``\textit{One thing I will point out is the incredibly steep technical cost of removing discriminatory language in software. For example, the fact that GitHub defaulted to the 'master' branch for many years has in turn created years of work for my team.}'' These remarks suggest that while many view language reform as a worthwhile step, they also recognize that large-scale terminology changes must be gradual and strategically implemented to minimize disruption.

\textbf{Gender and inclusion.}  
Our results also align with prior work \citep{zlotnick_2017}, which showed that women are more likely than men to encounter language or content that makes them feel unwelcome (25\% vs. 15\%) and to experience stereotyping (12\% vs. 2\%). In our study, female and non-binary participants expressed stronger agreement with the hypothesis that non-inclusive terminology can negatively affect inclusion and belonging. One participant noted, ``\textit{It erases my existence, so people do not associate women or neurodiverse people with technical excellence.}'' Such perspectives reinforce that language (even when used in a purely technical context) can contribute to feelings of exclusion for underrepresented groups. \nb{It should be noted that the gender-related impact differences observed in RQ2b do not survive Holm--Bonferroni correction and should be interpreted as preliminary, directional evidence rather than confirmed findings.}

\nb{\textbf{Selection effects and survivorship bias.}
The relatively modest differences observed between demographic groups (for example, between female and male respondents) may partly reflect selection effects within the developer population, including survivorship bias}~\citep{qiu2019going}. \nb{Developers who find non-inclusive language most harmful may disproportionately leave open-source communities, resulting in a sample that underrepresents those most negatively affected. This interpretation aligns with the finding that younger participants descriptively reported stronger negative effects on belonging, productivity, and self-esteem than older participants (a pattern that did not survive Holm--Bonferroni correction, but remains consistent with the survivorship-bias interpretation): those most affected may eventually leave the field or adapt, leaving a survivor population that appears less sensitive. Future research targeting recently disengaged contributors could help test this hypothesis.}

\textbf{Contextual interpretation and linguistic relativity.}  
A significant portion of respondents (76\%) who disagreed or remained neutral toward the hypothesis argued that the meaning of these terms depends primarily on their technical context. Many emphasized that words such as ``master'' or ``slave'' are used metaphorically in computing and are detached from social meaning. For example, one participant stated, ``\textit{As the grandson of a Black grandfather descended from slaves, I can happily say that this victimism over words is irrelevant to true professionals in the field.}'' Another remarked, ``\textit{The context in which these terms are used relates to software, not people.}''  

However, others pointed out that context does not entirely erase historical or cultural associations. One respondent explained, ``\textit{The understanding of most of these terms as offensive does not take into account context and basic linguistic principles.}'' Another acknowledged, ``\textit{Labeling terms such as master/slave out of context is a problem since they have a specific meaning in software.}'' These mixed views illustrate the ongoing tension between technical precision and social sensitivity in the use of language in professional settings.

\textbf{Summary and implications.}  
Taken together, our findings indicate that developers’ perceptions of non-inclusive terminology are shaped by a complex interplay of cultural background, linguistic familiarity, gender identity, and professional experience. For many, awareness of non-inclusive language arises primarily within Western or U.S.-based cultural frames, while others view such discourse as less relevant to their own local contexts. Some respondents perceive terminology reform as symbolic or even superficial, whereas others view it as a meaningful first step toward broader inclusion.  

Ultimately, the discussion suggests that efforts to promote inclusive language in software engineering should account for both cultural diversity and the pragmatic realities of software maintenance. Broad-based awareness campaigns, combined with gradual terminology updates and ongoing DEI initiatives, may offer the most effective path forward. This approach balances linguistic sensitivity with the need for technical stability and global inclusivity.

\section{Threats To Validity}\label{sec:threats_to_validity}

\textbf{Internal validity.}  
Internal validity addresses study design factors that could have influenced the results. Survey responses were analyzed statistically across multiple demographic characteristics, and significant differences were reported where present. Unobserved factors, such as the diversity climate of respondents’ teams or the broader culture of their development communities, may have also affected participants’ perceptions. These contextual variables were outside the scope of the current analysis and represent important directions for future research. Subsequent studies could investigate how organizational norms, communication practices, or diversity, equity, and inclusion (DEI) policies influence developers’ sensitivity to non-inclusive terminology.

\textbf{External validity.}  
External validity refers to the extent to which these findings can be generalized beyond the studied sample. The dataset includes 1,212 complete responses from active open-source software contributors, primarily developers involved in highly visible, community-oriented projects on GitHub. As a result, the findings may not fully capture the perspectives of developers in smaller open-source ecosystems, closed-source industrial environments, or geographically localized projects. Expanding future studies to include contributors from a broader range of development contexts could provide more comprehensive insights.

Another important consideration is potential self-selection bias. Voluntary participation in the survey may have resulted in a sample skewed toward individuals with greater interest in or awareness of language inclusivity issues. This limitation is common in perception-based online surveys. Future research could address this limitation by employing alternative sampling strategies, such as collaborations with professional organizations, targeted recruitment within companies, or controlled experimental designs, to capture a broader spectrum of perspectives.

In addition, the analysis was limited to a specific subset of terminology previously identified as non-inclusive in software artifacts. Therefore, the findings should not be generalized to other linguistic expressions or emerging terminology that were not examined in this study.

\textbf{Construct validity.}  

Construct validity refers to the extent to which the survey accurately measures the intended phenomena. \nb{One specific threat to construct validity in this study is potential instrument bias. Because the survey presented participants exclusively with terms previously identified as potentially non-inclusive, without including neutral or unproblematic control terms, respondents may have been inadvertently primed to consider non-inclusive interpretations of the presented vocabulary. Although this design choice was aligned with the study’s goal of examining perceptions of candidate non-inclusive terminology, it may have inflated agreement scores relative to what would be observed in a less prompted setting. Future studies could mitigate this threat by interspersing neutral control terms alongside candidate non-inclusive terms.}

As with most perception-based studies, the results reflect developers’ self-reported attitudes rather than direct observations of workplace behavior or performance. Although this approach yields valuable insights into individual perceptions, it does not quantify the actual impact of non-inclusive terminology on productivity, retention, or team cohesion. Future research could employ complementary methodologies, such as ethnographic fieldwork, repository content analysis, or longitudinal studies, to triangulate these perceptions with observable outcomes and strengthen the connection between theory and practice.

\section{Related Work}\label{sec:related_work}
\subsection{Team Diversity in Software Development}

Various aspects of team diversity influence professionals, teams, and their products, such as gender \citep{vasilescu2015gender,blincoe2019perceptions}, nationality, and spoken language \citep{daniel2013effects,ortu2017diverse}, experience \citep{chen2010effects}, knowledge \citep{liang2007effect}, and values \citep{liang2007effect}.

\citet{vasilescu2015gender} investigated the effects of gender and tenure diversity on the productivity and collaboration of GitHub teams. They discovered a positive correlation between both gender and tenure diversity and team performance, as measured by pull request acceptance rates.

\citet{blincoe2019perceptions} conducted a study to examine the impact of gender diversity on the mood and working atmosphere of software development teams. Their findings suggest that gender-diverse teams create a more pleasant and friendly atmosphere than male-dominated teams. However, they also encountered challenges such as gender discrimination and stereotypes. \nb{In a classroom-based case study, \citet{GARCIA2026112644} observed that an all-woman software engineering team demonstrated more help-seeking and leadership behaviors than mixed or male-majority teams, while men responded more slowly to communications; the study linked higher team engagement to improved conceptual understanding of software development. Building on the relationship between gender and participation in SE contexts, \citet{OBRIEN2025112225} developed gender-specific student personas to analyze learning needs and usability challenges in CS/SE education. Their findings suggest that design choices in educational tools may unintentionally disadvantage women due to differences in learning traits and problem-solving strategies.}

\citet{daniel2013effects} analyzed the effects of spoken language and nationality diversity across 357 SourceForge projects. They examined how three types of diversity (separation, variety, and disparity) influence community engagement and market success of open-source software projects. Their results indicate that variety diversity positively affects both outcomes, while separation diversity and disparity diversity have mixed effects depending on the project stage.

\citet{ortu2017diverse} carried out an empirical study to investigate the influence of gender and nationality diversity on the productivity and politeness of GitHub teams. They determined that higher gender diversity is associated with shorter bug-fixing times, increased productivity, and more positive sentiment, while higher nationality diversity is linked to lower politeness.

\citet{chen2010effects} examines how diversity in experience and interest affects group productivity and member withdrawal in online volunteer groups, using WikiProjects as a case study. They find that diversity in experience has an inverted U-shaped relationship with productivity and withdrawal, while diversity in interest has a positive linear relationship with both outcomes.

\citet{liang2007effect} examined the impact of knowledge diversity (KD) and value diversity (VD) on software team performance through task conflict and relationship conflict. Their findings indicate that KD positively affects performance by increasing task conflict, while VD negatively affects performance by increasing relationship conflict.\nb{More recently, \citet{torchiano2025impact} reported a moderate positive correlation between gender diversity and team outcomes in agile student projects, with no significant downside from nationality diversity. \citet{cha2024understanding} highlighted how accessibility barriers faced by blind and low-vision software professionals can limit their career mobility, emphasizing the need to treat disability inclusion as a core component of team diversity.}

In order to ensure team diversity and reap its benefits, it is crucial to understand the factors that may hinder it. In this paper, we focus on examining how the use of non-inclusive language can influence the productivity and well-being of individual developers, as well as the diversity of the teams they work within.


\subsection{Non-Inclusive Language in Software }

The study of non-inclusive language in software artifacts is in its incipient phases. Recent efforts, such as the comprehensive study by \citet{winchester2023harmful}, have brought attention to harmful terminology in computing. Their findings revealed that terms such as master/slave and blacklist/whitelist might carry negative connotations or implications for certain groups of people. \nb{Similarly, \citet{grabl2022scratch} examined how young Scratch programmers engage with socially relevant themes, including the Black Lives Matter movement and debates over terms like whitelist/blacklist. While the platform is primarily educational, their findings show that emerging programmers are exposed to and reflect broader discussions of inclusion in technology.} This observation aligns with the viewpoint of \citet{gilbert2022words} in their article, where they emphasized the importance of word choice in the computing field due to its potential to shape perception and inclusivity. Their stance also resonates with the “Words Matter” movement highlighted in their article, which focuses on problematic jargon in various domains.

Evidently, the discourse around non-inclusive terminologies is expanding. For instance, \citet{landau2020tech} reports the tech industry's introspection regarding the terms master and slave, especially in the backdrop of the Black Lives Matter movement. It sheds light on the longstanding usage of these terms in various technical contexts and emphasizes the ongoing debates concerning their replacement with more neutral terminologies.

Moreover, the use of such terminology extends beyond software to engineering education. \citet{danowitz2021assessing} conducted an insightful study on the effects of master/slave terminology in engineering education, highlighting the varying perceptions among students. Their findings underscore the importance of adopting more inclusive language to foster a sense of belonging.

Returning to the study by \citet{winchester2023harmful}, they noted that many organizations and companies have started changing their usage patterns of these problematic terms. Alongside their extensive database of terms classified as harmful in computing, they introduced a tool to detect and replace harmful computing-related terminology in documents. \nb{Building on these efforts, \citet{Win2023Hate} presented HaTe Detector, a web-based tool that integrates with GitHub to identify and suggest corrections for harmful terminology in markdown files. Their goal is to scale detection capabilities across repositories and promote inclusive language practices in online programming communities. Similarly, \citet{todd2024githubinclusifier} introduced the GitHubInclusifier, which detects non-inclusive language across multiple artifact types (including README files, PDFs, code, and comments) and proposes edits using large language models. Their tool can automatically open issues or commit fixes, aiming to support large-scale adoption of inclusive practices in open-source repositories.} 

Building on this foundation of identifying and rectifying non-inclusive language in software, our research focuses on a more subjective aspect-- the perception and impact of such language on developers. To the best of our knowledge, our work is the first to study developers' perceptions of the use of specific non-inclusive terms and the potential impact they can have on developer productivity, inclusion, well-being, and belonging. The closest related research to ours has primarily concentrated on several related aspects, such as toxic language in developer communications \citep{miller2022did}, methods for automatically detecting toxic or hostile language \citep{raman2020stress,sarker2020benchmark,cheriyan2021exploring}, the presence, content, and implementation of Codes of Conduct (CoCs) in open-source software (OSS) \citep{singh2021codes,tourani2017code}, their influence on the success of OSS \citep{coelho2017modern}, the participation of women in OSS projects \citep{singh2021codes}, and incivility in open-source code review discussions \citep{ferreira2021shut}.

\citet{miller2022did} investigated toxicity characteristics in open-source discussions on GitHub by analyzing a sample of 100 toxic issue comments. The findings revealed that toxicity in open-source platforms differs from other platforms in that it is more contextual, entitled, subtle, and passive-aggressive. Additionally, the study demonstrated that project members could also be sources of toxicity.

\citet{raman2020stress} explored the impact of toxic interactions on GitHub on stress and burnout experienced by open-source developers. The study showed that frequent demands for features and bug fixes, coupled with the occasionally aggressive tone of these demands, can contribute to unhealthy interactions. The paper outlined a path toward identifying, understanding, and mitigating such interactions by developing a machine learning classifier to recognize toxic comments in GitHub issues.

\citet{sarker2020benchmark} evaluated five toxicity detectors, including one specifically designed for the software engineering (SE) domain by \citet{raman2020stress}, on two extensive SE data sets of code review comments and Gitter messages. The study found that all detectors performed poorly on SE data, suggesting recommendations for improving their performance.

 \citet{cheriyan2021exploring} reported on a participatory study that compared the career motivations, strengths, and challenges of autistic and non-autistic university students. The findings indicated that autistic students face more barriers to employment than their non-autistic peers, and the study suggested ways to improve support for autistic students.

\citet{singh2021codes} examined the presence and effectiveness of codes of conduct (CoCs) in improving women's participation in Free and Open Source Software (F/OSS) communities. Their study analyzed 355 OSS websites for CoC presence and diversity elements, as well as five women-focused OSS forums for CoC usage and challenges. The results demonstrated that CoCs could positively impact women's participation, and the authors provided recommendations for better enforcement and reflection on the ethical underpinnings of CoCs as a tool to improve diversity and inclusion in OSS.

\citet{tourani2017code} conducted research on the role, scope, and influence of codes of conduct in open-source software (OSS) projects. They analyzed a variety of CoC templates and examples and discussed how these documents could foster diversity and inclusion within OSS communities.

\citet{coelho2017modern} surveyed maintainers of 104 deprecated GitHub repositories to explore the reasons behind the failure of modern open-source projects. The study identified nine factors that contribute to failure, which they categorized into three groups -- development team-related, development environment-related, and project characteristics-related. They also examined the relationship between specific maintenance practices and project success or failure, noting that development team-related reasons for failure included conflicts among developers, lack of interest, and lack of time.

\citet{ferreira2021shut} investigated the phenomenon of incivility in open-source code review discussions by analyzing 1,545 emails from the Linux Kernel Mailing List associated with rejected changes. The study identified various features, causes, and consequences of uncivil communication and provided insights for improving collaboration and communication in software engineering activities.

\nb{Recent work has also begun exploring how language bias in AI systems affects workplace communication and perceived inclusion. \citet{kadoma2024role} found that while stylistic suggestions from LLMs (e.g., hesitant vs. self-assured tones) did not directly change perceived inclusion, individuals who felt more included reported stronger ownership and agency over their writing, particularly among participants of minoritized genders. This highlights how linguistic alignment, even in mediated settings, can shape developers’ sense of control and belonging.}

Despite progress in understanding toxic language and CoCs in OSS communities, a gap exists regarding the impact of non-inclusive language on developers. Our study fills this gap by exploring developers' perceptions of such terms and their effects on productivity, inclusion, well-being, and belonging, contributing to the ongoing efforts to create healthier and more inclusive software development environments.

\section{Conclusions and Future Work}\label{sec:Conclusion}

This study explored software developers’ perceptions of non-inclusive terminology in software artifacts and its potential influence on their professional experiences and sense of inclusion. Based on 1,212 survey responses from active open-source contributors, we found that most participants were neutral or disagreed that the studied terms were non-inclusive. However, perceptions varied across demographics: women, non-binary, older, and U.S.-based respondents were more likely to view certain terms as non-inclusive, while others were more neutral. These results highlight the cultural and linguistic context underpinning perceptions of inclusivity in software language.

\nb{Perceived impacts showed directional differences by age and gender, though none of these survived Holm--Bonferroni multiple-comparison correction and should be treated as exploratory. Among participants who recognized the non-inclusive nature of the terms, younger respondents (18--38) descriptively reported stronger negative effects on belonging, productivity, and self-esteem compared to older respondents; female and other-gender participants also descriptively reported greater impacts than male respondents, especially regarding belonging and inclusion. In contrast, programming experience, occupation, and U.S./non-U.S. background showed no statistically significant differences in impact among those who agreed with the hypothesis. Together, these patterns suggest that age and gender identity may be stronger predictors of sensitivity to non-inclusive language than professional experience alone, though replication in larger samples is needed to confirm these trends.}

Our findings underscore that inclusive terminology is not only a matter of social responsibility but also one that can influence collaboration and team cohesion. As diverse and inclusive teams contribute to innovation and software quality, fostering awareness of inclusive communication can enhance both equity and effectiveness in the workplace. 

Future work should investigate these issues longitudinally and across more diverse populations and contexts, including industry environments. By providing empirical evidence on how developers perceive non-inclusive terminology, this study contributes to a growing understanding of how language and inclusion intersect in software engineering practice.

\section*{Acknowledgements}

This project was funded by the Deanship of Scientific Research (DSR) at King Abdulaziz University, Jeddah, Saudi Arabia, under grant no.\ (IPP: 1304-611-2025). The authors therefore acknowledge DSR with thanks for its technical and financial support.

\bibliographystyle{ACM-Reference-Format}
\bibliography{sample-base}

@article{ancona1992demography,
  title={Demography and design: Predictors of new product team performance},
  author={Ancona, Deborah Gladstein and Caldwell, David F},
  journal={Organization science},
  volume={3},
  number={3},
  pages={321--341},
  year={1992},
  publisher={INFORMS}
}

@article{bantel1989top,
  title={Top management and innovations in banking: Does the composition of the top team make a difference?},
  author={Bantel, Karen A and Jackson, Susan E},
  journal={Strategic management journal},
  volume={10},
  number={S1},
  pages={107--124},
  year={1989},
  publisher={Wiley Online Library}
}

@article{blincoe2019perceptions,
  title={Perceptions of gender diversity's impact on mood in software development teams},
  author={Blincoe, Kelly and Springer, Olga and Wrobel, Michal R},
  journal={Ieee Software},
  volume={36},
  number={5},
  pages={51--56},
  year={2019},
  publisher={IEEE}
}

@inproceedings{chen2010effects,
  title={The effects of diversity on group productivity and member withdrawal in online volunteer groups},
  author={Chen, Jilin and Ren, Yuqing and Riedl, John},
  booktitle={Proceedings of the SIGCHI conference on human factors in computing systems},
  pages={821--830},
  year={2010}
}

@inproceedings{coelho2017modern,
  title={Why modern open source projects fail},
  author={Coelho, Jailton and Valente, Marco Tulio},
  booktitle={Proceedings of the 2017 11th Joint meeting on foundations of software engineering},
  pages={186--196},
  year={2017}
}

@inproceedings{miller2022did,
  title={"Did you miss my comment or what?" understanding toxicity in open source discussions},
  author={Miller, Courtney and Cohen, Sophie and Klug, Daniel and Vasilescu, Bogdan and KaUstner, Christian},
  booktitle={Proceedings of the 44th International Conference on Software Engineering},
  pages={710--722},
  year={2022}
}

@article{ortu2017diverse,
  title={How diverse is your team? Investigating gender and nationality diversity in GitHub teams},
  author={Ortu, Marco and Destefanis, Giuseppe and Counsell, Steve and Swift, Stephen and Tonelli, Roberto and Marchesi, Michele},
  journal={Journal of Software Engineering Research and Development},
  volume={5},
  number={1},
  pages={1--18},
  year={2017},
  publisher={SpringerOpen}
}

@article{pelled1997demographic,
  title={Demographic diversity in work groups: An empirical assessment of linkages to intragroup conflict and performance},
  author={Pelled, L and Eisenhardt, K and Xin, K},
  journal={School of Business, University of Southern California},
  year={1997}
}

@inproceedings{qiu2019going,
  title={Going farther together: The impact of social capital on sustained participation in open source},
  author={Qiu, Huilian Sophie and Nolte, Alexander and Brown, Anita and Serebrenik, Alexander and Vasilescu, Bogdan},
  booktitle={2019 ieee/acm 41st international conference on software engineering (icse)},
  pages={688--699},
  year={2019},
  organization={IEEE}
}

@inproceedings{raman2020stress,
  title={Stress and burnout in open source: Toward finding, understanding, and mitigating unhealthy interactions},
  author={Raman, Naveen and Cao, Minxuan and Tsvetkov, Yulia and K{\"a}stner, Christian and Vasilescu, Bogdan},
  booktitle={Proceedings of the ACM/IEEE 42nd International Conference on Software Engineering: New Ideas and Emerging Results},
  pages={57--60},
  year={2020}
}

@inproceedings{sarker2020benchmark,
  title={A benchmark study of the contemporary toxicity detectors on software engineering interactions},
  author={Sarker, Jaydeb and Turzo, Asif Kamal and Bosu, Amiangshu},
  booktitle={2020 27th Asia-Pacific Software Engineering Conference (APSEC)},
  pages={218--227},
  year={2020},
  organization={IEEE}
}

@inproceedings{tourani2017code,
  title={Code of conduct in open source projects},
  author={Tourani, Parastou and Adams, Bram and Serebrenik, Alexander},
  booktitle={2017 IEEE 24th international conference on software analysis, evolution and reengineering (SANER)},
  pages={24--33},
  year={2017},
  organization={IEEE}
}

@inproceedings{vasilescu2015gender,
  title={Gender and tenure diversity in GitHub teams},
  author={Vasilescu, Bogdan and Posnett, Daryl and Ray, Baishakhi and van den Brand, Mark GJ and Serebrenik, Alexander and Devanbu, Premkumar and Filkov, Vladimir},
  booktitle={Proceedings of the 33rd annual ACM conference on human factors in computing systems},
  pages={3789--3798},
  year={2015}
}

@article{williams1998reilly,
  title={’Reilly. 1998. Demography and diversity in organizations: A review of 40 years of research},
  author={Williams, Katherine Y and Charles, AO},
  journal={Research in organizational behavior},
  volume={20},
  number={20},
  pages={77--140},
  year={1998}
}

@article{daniel2013effects,
  title={The effects of diversity in global, distributed collectives: A study of open source project success},
  author={Daniel, Sherae and Agarwal, Ritu and Stewart, Katherine J},
  journal={Information Systems Research},
  volume={24},
  number={2},
  pages={312--333},
  year={2013},
  publisher={INFORMS}
}

@article{eglash_2007, 
    title={Broken Metaphor: The Master-Slave Analogy in Technical Literature}, 
    volume={48}, 
    url={https://www.jstor.org/stable/40061475}, 
    number={2}, 
    journal={Technology and Culture}, 
    author={Eglash, Ron}, 
    year={2007}, 
    pages={360–369} 
}

@article{liang2007effect,
  title={Effect of team diversity on software project performance},
  author={Liang, Ting-Peng and Liu, Chih-Chung and Lin, Tse-Min and Lin, Binshan},
  journal={Industrial Management \& Data Systems},
  volume={107},
  number={5},
  pages={636--653},
  year={2007},
  publisher={Emerald Group Publishing Limited}
}

@inproceedings{torchiano2025impact,
  title={The Impact of Team Diversity in Agile Development Education},
  author={Torchiano, Marco and Coppola, Riccardo and Vetr{\`o}, Antonio and Musaj, Xhoi},
  booktitle={Proceedings of the 33rd ACM International Conference on the Foundations of Software Engineering},
  pages={1593--1601},
  year={2025}
}

@inproceedings{cha2024understanding,
  title={Understanding the career mobility of blind and low vision software professionals},
  author={Cha, Yoonha and Jackson, Victoria and Figueira, Isabela and Branham, Stacy Marie and Van der Hoek, Andr{\'e}},
  booktitle={Proceedings of the 2024 IEEE/ACM 17th International Conference on Cooperative and Human Aspects of Software Engineering},
  pages={170--181},
  year={2024}
}

@inproceedings{kadoma2024role,
  title={The role of inclusion, control, and ownership in workplace ai-mediated communication},
  author={Kadoma, Kowe and Aubin Le Quere, Marianne and Fu, Xiyu Jenny and Munsch, Christin and Metaxa, Dana{\"e} and Naaman, Mor},
  booktitle={Proceedings of the 2024 CHI Conference on Human Factors in Computing Systems},
  pages={1--10},
  year={2024}
}

@inproceedings{todd2024githubinclusifier,
  title={GitHubInclusifier: Finding and fixing non-inclusive language in GitHub Repositories},
  author={Todd, Liam and Grundy, John and Treude, Christoph},
  booktitle={Proceedings of the 2024 IEEE/ACM 46th International Conference on Software Engineering: Companion Proceedings},
  pages={89--93},
  year={2024}
}

@INPROCEEDINGS{Win2023Hate,
  author={Winchester, Hana and Al Haque, Ebtesam and Boyd, Alicia and Johnson, Brittany},
  booktitle={2023 IEEE Symposium on Visual Languages and Human-Centric Computing (VL/HCC)}, 
  title={HaTe Detector: A Tool for Detecting and Correcting Harmful Terminology in Computing Artifacts}, 
  year={2023},
  volume={},
  number={},
  pages={245-248},
  doi={10.1109/VL-HCC57772.2023.00043}}

@article{OBRIEN2025112225,
title = {Assessing gender bias in the software used in computer science and software engineering education},
journal = {Journal of Systems and Software},
volume = {219},
pages = {112225},
year = {2025},
issn = {0164-1212},
doi = {https://doi.org/10.1016/j.jss.2024.112225},
url = {https://www.sciencedirect.com/science/article/pii/S0164121224002693},
author = {Lyndsey O’Brien and Tanjila Kanij and John Grundy}
}

@INPROCEEDINGS{grabl2022scratch,
  author={Graßl, Isabella and Fraser, Gordon},
  booktitle={2022 IEEE/ACM 44th International Conference on Software Engineering: Software Engineering in Society (ICSE-SEIS)}, 
  title={SCRATCH as Social Network: Topic Modeling and Sentiment Analysis in SCRATCH Projects}, 
  year={2022},
  volume={},
  number={},
  pages={143-148},
  doi={10.1145/3510458.3513021}}

@article{GARCIA2026112644,
title = {A case study of gender and online team communication in Software Engineering Education},
journal = {Journal of Systems and Software},
volume = {231},
pages = {112644},
year = {2026},
issn = {0164-1212},
doi = {https://doi.org/10.1016/j.jss.2025.112644},
url = {https://www.sciencedirect.com/science/article/pii/S0164121225003139},
author = {Rita Garcia and Christoph Treude}
}

@article{singh2021codes,
  title={Codes of conduct in Open Source Software—for warm and fuzzy feelings or equality in community?},
  author={Singh, Vandana and Bongiovanni, Brice and Brandon, William},
  journal={Software Quality Journal},
  pages={1--40},
  year={2021},
  publisher={Springer}
}

@article{beecham2008motivation,
  title={Motivation in Software Engineering: A systematic literature review},
  author={Beecham, Sarah and Baddoo, Nathan and Hall, Tracy and Robinson, Hugh and Sharp, Helen},
  journal={Information and software technology},
  volume={50},
  number={9-10},
  pages={860--878},
  year={2008},
  publisher={Elsevier}
}

@article{cheriyan2021exploring,
  title={Exploring the career motivations, strengths, and challenges of autistic and non-autistic university students: Insights from a participatory study},
  author={Cheriyan, Chinnu and Shevchuk-Hill, Sergey and Riccio, Ariana and Vincent, Jonathan and Kapp, Steven K and Cage, Eilidh and Dwyer, Patrick and Kofner, Bella and Attwood, Helen and Gillespie-Lynch, Kristen},
  journal={Frontiers in Psychology},
  pages={4455},
  year={2021},
  publisher={Frontiers}
}

@article{ferreira2021shut,
  title={The" shut the f** k up" phenomenon: Characterizing incivility in open source code review discussions},
  author={Ferreira, Isabella and Cheng, Jinghui and Adams, Bram},
  journal={Proceedings of the ACM on Human-Computer Interaction},
  volume={5},
  number={CSCW2},
  pages={1--35},
  year={2021},
  publisher={ACM New York, NY, USA}
}

@misc{college_2020, 
    title={Gender Pronouns}, 
    url={https://springfield.edu/gender-pronouns/}, 
    journal={Springfield College}, 
    author={Canales, Katie}, 
    year={2020}
}

@misc{cosset_2020, 
    title={Replacing master in git}, 
    url={https://dev.to/damcosset/replacing-master-in-git-2jim}, 
    journal={DEV}, 
    author={Cosset, Damien}, 
    year={2020}, 
    month={Jun}
}

@misc{django_2014, 
    title={Replace occurrences of master/slave terminology with leader/follower}, 
    url={https://code.djangoproject.com/ticket/22667}, 
    journal={Django}, 
    author={Curella, Flavio}, 
    year={2014}, 
    month={May}
}

@misc{drupal_2014, 
    title={Replace "master/slave" terminology with "primary/replica"}, 
    url={https://www.drupal.org/project/drupal/issues/2275877}, 
    journal={Drupal}, 
    author={Feenstra, Bart}, 
    year={2014}, 
    month={May}
}

@misc{eurostat_2022, 
    title={ICT specialists in employment}, 
    url={https://ec.europa.eu/eurostat/statistics-explained/index.php?title=ICT_specialists_in_employment}, 
    journal={Eurostat}, 
    year={2022}, 
    month={Apr}
}

@misc{google_2022a, 
    title={Google Developer documentation Style Guide - Word list}, 
    url={https://developers.google.com/style/word-list#dummy-variabl}, 
    journal={Google Developer}, 
    year={2022}
}

@misc{google_2022b, 
    title={Google Developer documentation Style Guide - Write inclusive documentation}, 
    url={ https://developers.google.com/style/inclusive-documentation}, 
    year={2022}
}

@misc{greenblatt_2013, 
    title={The Racial History Of The 'Grandfather Clause'}, 
    url={https://www.npr.org/sections/codeswitch/2013/10/21/239081586/the-racial-history-of-the-grandfather-clause}, 
    journal={npr}, 
    author={Greenblatt, Alan}, 
    year={2013}, 
    month={Oct}
}

@misc{hansen_2017, 
    title={Ableist Language in Code: Sanity Check}, 
    url={https://gist.github.com/seanmhanson/fe370c2d8bd2b3228680e38899baf5cc#file-ableismsanitycheck-md}, 
    journal={GitHub}, 
    author={Hanson, Seán}, 
    year={2017}, 
    month={Apr}
}

@misc{hunter_2020, 
    title={Terms Like ‘Slave’ and ‘Master’ Finally Have Their Reckoning. It’s a Start.}, 
    url={https://builtin.com/software-engineering-perspectives/offensive-code-terminology-changes}, 
    journal={Built}, 
    author={Hunter, Tatum}, 
    year={2020}, 
    month={Jul}
}

@misc{inclusive_naming_initiative_2021, 
    title={Inclusive Naming Initiative}, 
    url={https://inclusivenaming.org/word-lists/tier-1/}, 
    journal={Inclusive Naming Initiative}, 
    year={2021}, 
}

@misc{landau_2020, 
    title={Tech Confronts Its Use of the Labels ‘Master’ and ‘Slave’}, 
    url={https://www.wired.com/story/tech-confronts-use-labels-master-slave/}, 
    journal={WIRED}, 
    author={Landau, Elizabeth}, 
    year={2020}, 
    month={Jul}
}

@misc{moore_1996, 
    title={Definition of Grandfathering in the Sense of Allowing a Preexisting Use to Continue Despite a Later Regulation}, 
    url={https://www.mtas.tennessee.edu/knowledgebase/definition-grandfathering-sense-allowing-preexisting-use-continue-despite-later}, 
    journal={University of Tennessee System}, 
    author={Moore, Todd}, 
    year={1996}, 
    month={May}
}

@misc{pwda_2022, 
    title={Language Guide}, 
    url={https://pwd.org.au/resources/language-guide/}, 
    journal={People with Disability Australia}, 
    year={2022}, 
}

@misc{python_2018, 
    title={Avoid master/slave terminology}, 
    url={https://bugs.python.org/issue34605}, 
    journal={Python Software Foundation}, 
    author={Vstinner}, 
    year={2018}, 
    month={Aug}
}

@misc{riley_2019, 
    title={Words Matter: Why We Should Put an End to “Grandfathering”}, 
    url={https://medium.com/@nriley/words-matter-why-we-should-put-an-end-to-grandfathering-8b19efe08b6a}, 
    journal={Medium}, 
    author={Riley, Nancy}, 
    year={2019}, 
    month={May}
}

@misc{shankland_2020, 
    title={Twitter engineers replacing racially loaded tech terms like 'master,' 'slave'}, 
    url={https://www.cnet.com/culture/twitter-engineers-replace-racially-loaded-tech-terms-like-master-slave/}, 
    journal={CENT}, 
    author={Shankland, Stephen}, 
    year={2020}, 
    month={Jul}
}

@misc{tobia_2016, 
    title={Everything You Ever Wanted to Know About Gender-Neutral Pronouns}, 
    url={https://time.com/4327915/gender-neutral-pronouns/}, 
    journal={TIME}, 
    author={Tobia, Jacob}, 
    year={2016}, 
    month={May}
}

@misc{stack_overflow_2021, 
    title={2021 Developer Survey}, 
    url={https://insights.stackoverflow.com/survey/2021}, 
    journal={Stack Overflow}, 
    year={2021}, 
}

@misc{us_bureau_labor_statistics_2021, 
    title={  Labor Force Statistics from the Current Population Survey}, 
    url={https://www.bls.gov/cps/cpsaat11.htm}, 
    journal={U.S. Bureau of Labor Statistics}, 
    year={2021}, 
}

@misc{us_census_2021, 
    title={U.S. Census Bureau QuickFacts: United States}, 
    url={https://www.census.gov/quickfacts/fact/table/US/PST045221}, 
    journal={United States Census Bureau}, 
    year={2020}
}

@misc{zlotnick_2017, 
    title={Github Open Source Survey 2017}, 
    url={https://opensourcesurvey.org/2017/}, 
    journal={Github Open Source Survey}, 
    year={2017}, 
}

@misc{replication_package,
    title={Replication Package: The Perception and Impact of Non-inclusive Language in Software Artifacts},
    author={Tayeb, Ahmad J. and Alahmadi, Mohammad D.},
    year={2026},
    publisher={Zenodo},
    doi={10.5281/zenodo.21098146},
    url={https://doi.org/10.5281/zenodo.21098146}

}

@inproceedings{winchester2023harmful,
  title={Harmful Terms in Computing: Towards Widespread Detection and Correction},
  author={Winchester, Hana and Boyd, Alicia E and Johnson, Brittany},
  booktitle={2023 IEEE/ACM 45th International Conference on Software Engineering: Software Engineering in Society (ICSE-SEIS)},
  pages={132--137},
  year={2023},
  organization={IEEE}
}

@article{gilbert2022words,
  title={Words matter},
  author={Gilbert, Juan E and Ludi, Stephanie and Patterson, David A and Smith, Lisa M},
  journal={Communications of the ACM},
  volume={65},
  number={7},
  pages={36--36},
  year={2022},
  publisher={ACM New York, NY, USA}
}

@article{landau2020tech,
  title={Tech confronts its use of the labels ‘master’and ‘slave’},
  author={Landau, Elizabeth},
  journal={WIRED. July},
  year={2020}
}

@inproceedings{danowitz2021assessing,
  title={Assessing the Effects of Master Slave Terminology on Inclusivity in Engineering Education},
  author={Danowitz, Andrew and Asfaw, Amman Fasil and Benson, Bridget and Hummel, Paul and McKell, K Clay},
  booktitle={2021 CoNECD},
  year={2021}
}

@misc{github_octoverse_2025,
  author       = {{GitHub}},
  title        = {The State of the Octoverse},
  year         = {2025},
  howpublished = {\url{https://octoverse.github.com/}},
  note         = {Accessed: 2026-03-23}
}

@misc{stackoverflow_survey_kaggle_2024,
  author       = {{Stack Overflow}},
  title        = {Stack Overflow Annual Developer Survey 2011--2024},
  year         = {2024},
  howpublished = {\url{https://www.kaggle.com/datasets/joebeachcapital/stack-overflow-annual-developer-survey-2024}},
  note         = {Kaggle dataset. Accessed: 2026-03-23}
}

\end{document}